\documentclass[aps, prd, twocolumn, tightenlines, notitlepage, superscriptaddress, nofootinbib, preprintnumbers, floatfix, showkeys,10pt]{revtex4-2}

\usepackage[normalem]{ulem}
\usepackage{amssymb}
\usepackage{amsmath}
\usepackage{graphicx}
\usepackage{url}
\usepackage{ulem}
\usepackage[utf8]{inputenc}
\usepackage{slashed}
\usepackage{lipsum}
\usepackage[T1]{fontenc}
\usepackage{appendix}
\usepackage{csquotes}
\usepackage[x11names,table]{xcolor}
\usepackage{comment}
\usepackage[export]{adjustbox}
\usepackage{textgreek}
\usepackage{caption, subcaption}
\usepackage{ragged2e} 
\DeclareCaptionJustification{justified}{\justifying}
\usepackage{booktabs}
\usepackage{orcidlink}
\usepackage{hhline}

\usepackage{hyperref}

\hypersetup{colorlinks,citecolor= nicered,linkcolor= blue}
\definecolor{nicered}{rgb}{0.7,0.1,0.1}
\definecolor{nicegreen}{rgb}{0.1,0.5,0.1}

\def\cevns{CE\textnu NS}
\def\eves{E\textnu ES}

\def\d{\mathrm{d}}

\definecolor{amber}{rgb}{1.0, 0.49, 0.0}
\definecolor{yellow-green}{rgb}{0.6, 0.8, 0.2}
\definecolor{gray(x11gray)}{rgb}{0.75, 0.75, 0.75}
\definecolor{lightsalmon}{rgb}{1.0, 0.63, 0.48}

\definecolor{myblue}{cmyk}{0.65, 0.37, 0.0, 0.19}

\definecolor{blue(ncs)}{rgb}{0.0, 0.53, 0.74}

\usepackage{multirow}

\AtBeginDocument{\hypersetup{citecolor=myblue,linkcolor=myblue,urlcolor=myblue}}

\begin{document}

\title{{\Large 
Searches for heavy neutral lepton decays\\ at spallation neutron sources}}

\author{Valentina De Romeri~\orcidlink{0000-0003-3585-7437}}\email{deromeri@ific.uv.es}
\affiliation{Instituto de F\'{i}sica Corpuscular (IFIC), CSIC‐Universitat de Val\'encia, E-46980 Valencia, Spain}

\author{A. Galindo-Uribarri~\orcidlink{0000-0001-7450-404X}}\email{aguman@gmail.com}\affiliation{Department of Physics and Astronomy, University of Tennessee, Knoxville, TN 37996, USA}

\author{Ana Martín-Galán~\orcidlink{0000-0001-6628-1851} }\email{ana.martin@ific.uv.es}
\affiliation{Instituto de F\'{i}sica Corpuscular (IFIC), CSIC‐Universitat de Val\'encia, E-46980 Valencia, Spain}
\author{Víctor Martín Lozano~\orcidlink{0000-0002-9601-0347}}\email{victor.lozano@ific.uv.es}
\affiliation{Instituto de F\'{i}sica Corpuscular (IFIC), CSIC‐Universitat de Val\'encia, E-46980 Valencia, Spain}
\affiliation{Departament de F\'isica Teòrica, Universitat de Val\`{e}ncia, 46100 Burjassot, Spain}
\author{G. Sanchez Garcia~\orcidlink{0000-0003-1830-2325}}%
\email{g.sanchez@ciencias.unam.mx}%
\affiliation{Departamento de F\'{i}sica, Facultad de Ciencias, Universidad Nacional Aut\'onoma de M\'exico,
Apartado Postal 70-542, Ciudad de M\'exico 04510, M\'exico}%

\begin{abstract}
Spallation neutron sources provide intense neutrino fluxes from pion and muon decay at rest, with energies in the few tens of MeV range. Experiments such as COHERENT exploit these fluxes to detect neutrinos via coherent elastic neutrino–nucleus scattering (\cevns). However, these facilities also offer a unique opportunity to produce and probe light, secluded, or weakly-coupled particles.
In this work, we investigate the sensitivity of the Spallation Neutron Source at the Oak Ridge National Laboratory to heavy neutral leptons (HNLs) in the MeV-GeV mass range, as a case study. We consider HNL production in pion and muon decays at rest, followed by their decay into visible Standard Model particles within the detector volume. We analyze a range of current and proposed COHERENT detectors and evaluate their sensitivity to HNL mixing with muon and electron neutrinos, as well as to scenarios with mixed mixing. We find that existing detectors can set meaningful constraints, particularly for muon-flavor mixing, while future ton-scale realizations can probe previously unexplored regions of the parameter space. We further discuss prospects at other relevant spallation source facilities and comment on the complementarity of their projected sensitivities. Our results demonstrate that \cevns~experiments at spallation neutron sources provide a powerful, complementary avenue for new physics searches beyond their primary role as neutrino detectors.
\end{abstract}

\maketitle

\section{Introduction}

Coherent elastic neutrino–nucleus scattering (\cevns) has recently emerged as a powerful and rapidly developing channel for neutrino detection via the neutral current~\cite{Abdullah:2022zue}. Its first observation~\cite{COHERENT:2017ipa,COHERENT:2021xmm} occurred decades after its theoretical prediction~\cite{Freedman:1973yd}, primarily due to the experimental challenges associated with measuring the tiny nuclear recoils involved. While the process itself has opened new opportunities for compact neutrino detectors, its realization has more broadly demonstrated the potential of low-energy, high-intensity neutrino sources as precision probes of fundamental physics. In particular, the experimental program built around the Spallation Neutron Source (SNS) at Oak Ridge National Laboratory has shown that such facilities can access a wide range of Standard Model (SM) tests as well as searches for physics beyond the Standard Model (BSM) (see, e.g.,~\cite{Cadeddu:2020lky,DeRomeri:2022twg,Breso-Pla:2025cul}).

Neutrinos produced at spallation sources arise predominantly from pion decay at rest, followed by muon decay, yielding fluxes with energies in the few tens of MeV range. Compared to reactor or solar neutrinos, these higher energies enhance sensitivity to nuclear structure effects~\cite{Papoulias:2019lfi,DeRomeri:2026dac,AtzoriCorona:2026wbu} and allow for a rich phenomenology. Crucially, the same production mechanisms that generate the neutrino flux can also yield new, weakly coupled particles in the MeV mass range, so that detectors built to measure \cevns~are simultaneously exposed to a potential dark sector. The physics potential of meson and muon decays at spallation facilities as probes of BSM physics has long been recognized, and includes, for instance, the production of dark-sector states via neutral pion decays, such as sub-GeV dark matter~\cite{deNiverville:2015mwa,Dutta:2019nbn,Dutta:2020vop,COHERENT:2019kwz,COHERENT:2021pvd,CCM:2021leg,CCM:2021yzc,COHERENT:2022pli,COHERENT:2023sol,AristizabalSierra:2026jgp}, or other exotic states~\cite{Ge:2017mcq,Calabrese:2022mnp,DeRomeri:2023cjt,Ema:2023buz}, also through upscattering~\cite{Brdar:2018qqj,Chang:2020jwl,Candela:2023rvt,Candela:2024ljb,Feng:2026jsi}. 
In this context, existing experimental efforts originally designed for \cevns~measurements, such as the COHERENT program, can be naturally reinterpreted as multipurpose detectors for new physics searches.

The growing \cevns~experimental landscape, which at present includes measurements with CsI, LAr~\cite{COHERENT:2020iec}, Ge~\cite{Adamski:2024yqt,COHERENT:2026yje} and NaI~\cite{COHERENT:2026ewu} targets at the SNS, as well as reactor~\cite{Colaresi:2022obx,Ackermann:2025obx} and dark matter direct detection experiments~\cite{XENON:2024ijk,PandaX:2024muv,LZ:2025igz,XENON:2026ydt}, has demonstrated the versatility of these setups. While all these measurements enable precision SM tests, also in combination with the neutrino-electron scattering channel~\cite{AtzoriCorona:2025xwr}, they also provide sensitivity to a broad class of new physics scenarios, including light and heavy mediators (see, e.g.,~\cite{DeRomeri:2026prc,Lozano:2025ekx}). This dual role highlights the importance of developing complementary search strategies that go beyond the \cevns~signal itself. 

In this work, we explore the sensitivity of \cevns~detectors at spallation neutron source facilities, in particular the SNS, to heavy neutral leptons (HNLs), focusing on signatures that do not rely on nuclear scattering. Gauge-singlet fermions like HNLs are well-motivated extensions of the SM and arise naturally in frameworks addressing neutrino masses, such as the \textit{seesaw} mechanism~\cite{Schechter:1980gr,Yanagida:1980xy,Minkowski:1977sc}. After electroweak symmetry breaking, HNLs interact with SM particles through mixing with active neutrinos, leading to a rich phenomenology that depends strongly on their mass~\cite{Abdullahi:2022jlv}. While very light sterile neutrinos are tightly constrained by oscillation and cosmological data~\cite{Abazajian:2012ys,Gariazzo:2015rra}, HNLs in the MeV-GeV range remain particularly appealing, as they can be probed in beam-dump~\cite{SHiP:2018xqw,Batell:2020vqn} and collider experiments~\cite{ATLAS:2025qbs,Cottin:2018nms,Drewes:2019fou,Bondarenko:2019tss}, as well as at intense neutrino sources~\cite{Ballett:2019bgd,Berryman:2019dme}. Additionally, the presence of HNLs can also induce deviations from unitarity in the leptonic mixing matrix, possibly within reach of \cevns~experiments~\cite{Miranda:2020syh,CentellesChulia:2025jir}.

The production and detection of HNLs in pion and muon decays at rest has been previously investigated~\cite{Ema:2023buz}, including sensitivity projections for future COHERENT detectors~\cite{Hostert:2025ffy} and for a ton-scale hydrocarbon scintillator
detector~\cite{PROSPECT:2026jsl}. Here, we build upon these studies and extend them by considering HNL production at the SNS and their subsequent decays within the detector volume, leading to observable signals -- displaced or prompt energy depositions from the HNL decay products -- that are kinematically and topologically distinct from the nuclear recoils characteristic of \cevns. We analyze a set of current and under deployment COHERENT detectors~\cite{COHERENT:2026ewu}, like the NaI, LAr, and D$_2$O targets, and evaluate their sensitivity to HNL mixing with both muon and electron neutrinos, as well as mixed scenarios. We show that despite their relatively small size (compared to other neutrino experiments), existing detectors can already set competitive constraints, particularly due to the sizable muon component in the SNS flux. We then consider proposed upgrades and future subsystems of the COHERENT program and demonstrate their potential to probe unexplored regions of parameter space or provide complementary to dedicated beam-dump and collider searches.

Although our analysis focuses on the SNS, the same strategy can be readily extended to other spallation neutron source facilities, where \cevns~detectors are already running or planned to be deployed soon, such as the Lujan facility at the
Los Alamos Neutron Science Center (LANSCE)~\cite{CCM:2021leg}, the European Spallation Source (ESS)~\cite{Baxter:2019mcx}, the Japan Proton Accelerator Research Complex (J-PARC)~\cite{Collar:2025sle}, and the China Spallation Neutron Source (CSNS)~\cite{Su:2023klh}. We therefore present projected sensitivities for some of these facilities, and discuss how they depend on key experimental parameters, including beam power, number of protons on target, and detector assumptions.

Our paper is organized as follows. In Sec.~\ref{sec:theory}, we present the theoretical framework for an SM extension with a single HNL. In Sec.~\ref{sec:HNLs_SNS}, we discuss the relevant HNL production and decay channels at spallation source facilities, focusing on the COHERENT program at the SNS, as well as the proposed \cevns~experiments at the ESS, J-PARC, and CSNS. Despite our analysis being purely phenomenological, we also comment on relevant experimental details of these searches, and on how the resulting sensitivities could be improved. 
We present our results in Sec.~\ref{sec:results}, and draw our conclusions in Sec.~\ref{sec:conclusions}.

\section{Theoretical framework}
\label{sec:theory}

The observation of neutrino oscillations has provided clear evidence of their non-zero masses~\cite{deSalas:2020pgw,Esteban:2024eli,Capozzi:2025wyn}.  However, the SM cannot accommodate neutrino mass terms in any of its usual forms: a Dirac mass would require right-handed neutrino fields, which have not been observed as SM gauge singlets, while a Majorana mass for the active neutrinos is forbidden by gauge invariance. Generating neutrino masses therefore requires BSM physics.

One of the simplest realizations consists of adding one or more gauge-singlet fermions, commonly referred to as HNLs or sterile neutrinos, that mix with the SM neutrinos.
While the number of HNLs is in principle unrestricted, in this work we assume that only one HNL, $N$, is kinematically accessible at the energies relevant for our analysis, and that it mixes with the active flavour states as
\begin{align}
    \nu_\ell = U_{\ell N} N + \sum_ {i=1}^{3}U_{\ell i} \nu_i,
\end{align}
where $\ell=e,\mu,\tau$ is the flavor index, $i$ runs over the three light mass eigenstates, and $U_{\ell N}$ parametrizes the mixing between the active flavor $\ell$ and the heavy state $N$.

In this minimal extension, the charged- and neutral-current interactions  between $N$ and the SM leptons are described by the effective Lagrangian
\begin{align}
\label{eq:HNL_lagrangian}
    \mathcal{L} \supseteq\, & -\dfrac{g}{\sqrt{2}}\, W_\mu^+ U_{\ell N}^{*}\, \overline{N}\,  \gamma^\mu P_L \ell \nonumber \\
    & -\dfrac{g}{2\cos\theta_W}\,  Z_\mu U_{\ell N}^{*}\, \overline{N} \, \gamma^\mu P_L \nu_\ell \;+\; \mathrm{h.c.},
\end{align}
where $g$ is the $SU(2)$ gauge coupling and $\theta_W$ is the weak mixing angle. The smallness of the mixing element $U_{\ell N}$, suppresses the interactions of the HNL with the SM fields relative to a typical weak-interaction process of the same order\footnote{In most experimental setups, the combined production and subsequent decay of the HNL scales as $|U_{\alpha N}|^4$, i.e.\ as the product of a production rate and a decay rate, each individually suppressed by $|U_{\alpha N}|^2$.}.
As a result, HNLs are generically expected to be long-lived, placing them in the broader class of long-lived particles (LLPs).
 
In full generality, more than one entry $U_{\ell N}$ of the mixing matrix can be phenomenologically relevant at low energies~\cite{Ballett:2019bgd,Abada:2022wvh}. We therefore consider, depending on the scenario, mixing with either a single flavor or with two flavors simultaneously, deriving constraints on $|U_{\alpha N}|^2$ and on $|U_{\alpha N} U_{\beta N}|$ (with $\alpha \neq \beta$), respectively.

Throughout this work we assume $N$ to be a Dirac fermion. This choice is motivated by low-scale seesaw constructions, such as the inverse~\cite{Mohapatra:1986bd,GONZALEZGARCIA1989360} and linear~\cite{Akhmedov:1995ip,Akhmedov:1995vm,Malinsky:2005bi} seesaw mechanisms, in which the HNL naturally arises as part of a pseudo-Dirac pair with a parametrically small lepton-number-violating (Majorana) mass splitting.
In this limit, the phenomenology of $N$ is governed by lepton-number-conserving interactions, and the Dirac approximation is well justified. For a purely Majorana HNL, production and decay rates receive additional lepton-number-violating contributions; existing literature indicates that the relevant rates are typically a factor of two larger than in the Dirac case (see, e.g.,~\cite{Atre:2009rg,Bondarenko:2018ptm,Helo:2010cw}).

\section{Heavy Neutral Leptons at spallation neutron sources}
\label{sec:HNLs_SNS}

The Spallation Neutron Source at Oak Ridge National Laboratory is a pulsed-beam facility where protons with $\sim 1.3$ GeV kinetic energy impinge on a fixed mercury target at 60 Hz with pulse widths as short as 400 ns wide~\cite{Haines:2014kna}. From this interaction, high-energy neutrons and other particles are produced, with charged and neutral pions being the dominant byproducts at this energy regime. Essentially all of the negatively charged pions are captured inside the target before decaying, while positively charged pions are stopped within the target and decay at rest (DAR) in approximately 99\% of cases~\cite{COHERENT:2021yvp}.
Within the SM, these pions decay into muons and neutrinos, with the muons subsequently decaying into positrons, electron neutrinos, and muon antineutrinos. The dominant SM neutrino production modes at the SNS are therefore the two-body pion decay followed by the three-body muon decay,
\begin{align}
    \pi^+ &\to \mu^+ \, \nu_\mu, \\
    \mu^+ &\to e^+ \, \nu_e \, \bar \nu_\mu \, .
\end{align}
Consequently, the resulting neutrino flux exhibits a prompt component from pion decay and a delayed component from muon decay.

The pulsed structure of the beam, together with the timing resolution of COHERENT-class detectors provides enough information to disentangle the prompt and delayed flux contributions~\cite{COHERENT:2021yvp}. If a heavy, long-lived new-physics state is produced in either the pion or muon decay, its arrival time at the detector will generally be delayed relative to the corresponding SM neutrino produced in the same decay, since a massive particle travels slower than a massless one for a given momentum. For a HNL state of mass $m_X$ and momentum $p$ produced at a baseline distance $L$ from the source, this delay is approximately $\delta t \approx L\,\frac{m_X^2}{2 p^2}$~\cite{Hostert:2025ffy}.
The delay therefore grows with the baseline and with the square of the HNL mass, while being suppressed for higher-momentum (more relativistic) states. Provided the detector has sufficient timing resolution, this mass- and momentum-dependent delay offers a handle to both identify the parent decay channel and discriminate the HNL signal from the SM prompt/delayed neutrino fluxes and from backgrounds.\\

In the following subsections, we discuss HNL production, decay and its detection at COHERENT suite of experiments at the SNS, as well as a brief comment on prospects at other spallation-source facilities.

\subsection{Production and decay of HNLs via mixing}

Given the interactions in Eq.~\eqref{eq:HNL_lagrangian}, HNLs can be produced at the SNS via the pion decay-at-rest channel $\pi^+ \to \ell^+ \, N$, as well as via the muon decay modes $\mu^+ \to e^+ \, \nu_e \, \bar N$ and $\mu^+ \to e^+ \, \bar\nu_\mu \, N$, depending on which HNL-neutrino mixing is active and on the kinematic accessibility set by the HNL mass $m_N$.

In pion decay, two channels involving HNLs can in principle contribute. If the kinematic condition $m_\pi > m_\mu + m_N$ is satisfied, the channel $\pi^+ \to \mu^+ \, N$ is allowed, and, additionally, if $m_\pi > m_e + m_N$, the channel $\pi^+ \to e^+ \, N$ is also open.

The total decay width for the two-body process in the pion rest frame is given by~\cite{Ema:2023buz}
\begin{align}
    \Gamma_{\pi \rightarrow \ell N}&=
     \dfrac{G_F^2 f_\pi^2 |V_{ud}|^2 |U_{\ell N}|^2}{8 \pi m_\pi^3} \nonumber\\
    &\times \left[m_\pi^2(m_\ell^2+m_N^2) - (m_\ell^2+m_N^2)^2 + 4 m_\ell^2m_N^2 \right] \\
    &\times \sqrt{m_\pi^4-2(m_\ell^2+m_N^2)m_\pi^2 + (m_\ell^2-m_N^2)^2}\, , \nonumber
\end{align}
where $G_F$ is the Fermi constant, $f_\pi$ is the pion decay constant, and $V_{ud}$ is the relevant CKM element for the up-down mixing. Since pions decay at rest, the angular distribution of the decay products is isotropic and no boost is required to obtain the resulting energy spectrum in the lab frame.
The muon produced in pion decay can subsequently decay into an HNL through two channels.
If $m_N < m_\mu - m_e$, the HNL can be produced via $\mu^+ \to \bar N \, e^+ \, \nu_e$, governed by the mixing $U_{\mu N}$, or via $\mu^+ \to e^+ \, \bar\nu_\mu \, N$, governed by $U_{eN}$.

The total decay rate can be obtained as
\begin{equation}
    \Gamma = \int_{m_N}^{(m_\mu^2 + m_N^2 ) / (2m_\mu)} \left[\dfrac{\mathrm{d}\Gamma}{\mathrm{d}E_N}\right] \mathrm{d}E_N\, ,
\end{equation}
where the differential decay rates for the two processes are
\begin{align}\label{decay_width_muon_case_biblio}
    \frac{d \Gamma (\mu \to e \bar \nu_e N)}{d E_N} &= \frac{G_F^2 |U_{\mu N}|^2}{12 \pi^3} \left( 3 E_N(m_\mu^2 + m_N^2) \right. \nonumber\\
    &- 4 m_\mu E_N^2 - \left. m_\mu m_N^2 \right) \sqrt{E_N^2 - m_N^2},
\end{align}
\begin{align}
    \dfrac{\mathrm{d}\Gamma(\mu \rightarrow \bar N e \nu_\mu)}{\mathrm{d}E_N} &= \frac{G_F^2 |U_{eN}|^2}{2 \pi^3} E_N \left( m_\mu^2 + m_N^2 - 2 m_\mu E_N \right) \nonumber\\
    &\times \sqrt{E_N^2 - m_N^2} \, ,
\end{align}
and we have taken $m_e \to 0$, given the energy scale of the process.
Given the nature of the HNL production channels considered, the parameter space accessible at these facilities is kinematically constrained by the pion mass, restricting our sensitivity to $m_N \lesssim 139$ MeV.

The relevant HNL decay channels at spallation neutron sources proceed via both neutral- and charged-current interactions. Through neutral currents, HNLs can decay as $N \to \nu_\mu \, e^- e^+$ when muon mixing is active, and as $N \to \nu_e \, e^- e^+$ when electron mixing is active. Additionally, a purely invisible channel $N \to \nu_\ell \nu_\alpha \bar \nu_\alpha$ is present independently of which mixing is active. Charged-current contributions are present only for electron mixing, via $N \to \nu_e e^- e^+$, and interfere with the corresponding neutral-current amplitude.

Still in the limit $m_e \rightarrow 0$, the partial decay widths are given by~\cite{Ema:2023buz}
\begin{align}
    \Gamma(N \rightarrow \nu_\mu e^- e^+) &= \dfrac{G_F^2 |U_{\mu N}|^2}{768\pi^3} m_N^5 \nonumber \\ 
    &\times \left(1 - 4\sin^2\theta_W + 8\sin^4\theta_W\right)\, ,
\end{align}
\begin{align}
    \Gamma(N \rightarrow \nu_e e^- e^+) &= \dfrac{G_F^2 |U_{e N}|^2}{768\pi^3} m_N^5 \nonumber \\
    &\times \left(1 + 4\sin^2\theta_W + 8\sin^4\theta_W\right)\, ,
\end{align}
\begin{equation}
    \Gamma(N \rightarrow \nu_\ell \nu_\alpha \overline{\nu}_\alpha) = \dfrac{G_F^2 |U_{\ell N}|^2}{768\pi^3} m_N^5\, ,
\end{equation}
where the last expression applies to each of the invisible decay modes considered. The corresponding expressions for the charge-conjugate (anti-HNL) processes are identical in all channels. While all three channels contribute to the total decay width of the HNL in the mass range of interest here, only the first two, which yield an electron-positron pair, can give rise to a visible signal inside the \cevns~detectors. 

\begin{table}
\renewcommand{\arraystretch}{1.3}
\begin{tabular}{|l|c|c|c|}
\toprule
\bf Detector & Mass {[}kg{]} & Size {[}cm{]} & Baseline {[}m{]} \\ \toprule \hline
\cellcolor{yellow-green!20}NaI v1~\cite{COHERENT:2026ewu} & \cellcolor{yellow-green!20}2400 & \cellcolor{yellow-green!20}87 & \cellcolor{yellow-green!20}22 \\ \hline
\cellcolor{yellow-green!20}NaI v2~\cite{COHERENT:2026ewu} &  \cellcolor{yellow-green!20}3500 & \cellcolor{yellow-green!20}97 & \cellcolor{yellow-green!20}22 \\ \hline
\cellcolor{yellow-green!20}LAr~\cite{COHERENT:2026ewu} &  \cellcolor{yellow-green!20}475 & \cellcolor{yellow-green!20}70 & \cellcolor{yellow-green!20}27.5 \\ \hline
\rowcolor{yellow-green!20}D$_2$O~\cite{COHERENT:2026ewu} & 550 & 80 & 19.3 \\ \hline
\cellcolor{lightsalmon!20}D$_2$O+H$_2$O~\cite{COHERENT:2026ewu} & \cellcolor{lightsalmon!20}1000 & \cellcolor{lightsalmon!20}102 & \cellcolor{lightsalmon!20}19.3 \\ \hline
\cellcolor{lightsalmon!20}LAr-750~\cite{COHERENT:2026ewu,COHERENT:2023sol} & \cellcolor{lightsalmon!20}610 & \cellcolor{lightsalmon!20}76 & \cellcolor{lightsalmon!20}27.5 \\ \hline
\cellcolor{lightsalmon!20}LAr-10t~\cite{COHERENT:2021pvd,COHERENT:2023sol} & \cellcolor{lightsalmon!20}10000 & \cellcolor{lightsalmon!20}192 & \cellcolor{lightsalmon!20}27.5 \\ \hline
\bottomrule
\end{tabular}
\caption{COHERENT detectors at the SNS and corresponding parameters considered in this work. For each detector we list the detector mass, linear detector size, and baseline distance $L$ from the SNS target. Detectors highlighted in green correspond to current experiments, while those in orange correspond to proposed future detectors. In all cases, we assume a number of protons on target $N^{\mathrm{POT}} = 1.728\times 10^{23}/$yr and an average number of $\pi^+$ produced per POT $c_{\pi^+}=0.11$.}
\label{tab:detectors}
\end{table}

\subsection{Detection of HNLs}
\label{subsec:det}

Since no experimental data are available for this specific search, we rely on simulated data for our analysis. We first simulate the flux for each of the outgoing particles from pion decay. Since pions are produced nearly at rest at the SNS, the HNL inherits only a mild boost, $\gamma \lesssim \mathcal{O}(\text{few})$, set by the small available phase space $m_\pi - m_\mu \simeq 34\,\text{MeV}$ for the two-body channel $\pi^+ \to \mu^+ N$, or by the proximity of $m_N$ to $m_\pi$ for $\pi^+ \to e^+ N$. Consequently, for the parameter space of interest and given the half life of the charged pions, such pion decays can be treated as occurring at rest and the subsequent HNL flux can be safely approximated as isotropic. The resulting four-momenta are used to construct the normalized energy spectra of the final-state particles. Then, the rescaled energy spectrum of HNLs produced via pion decay is given by
\begin{align}
    \label{eq:HNL_prod}
    \left.
    \dfrac{\d N^{\rm{HNL}}}{\d E}\right|^{\rm{prod}} = N^{\rm{POT}} \,  c_{\pi^+} 
    \, 
    \mathcal{B}(\pi^+ \to N\, e / \mu)
    \left.
    \dfrac{\d N^{\rm{HNL}}}{\d E}\right|^{\rm{norm=1}}\, ,
\end{align}
where $N^{\rm{POT}}$ is the total number of protons on target (POT) accumulated over the data-taking period, $c_{\pi^+}$ denotes the average number of $\pi^+$ produced per POT, $\mathcal{B}$ is the branching ratio for a given production channel ($\ell = e,\mu$) and $\left. \dfrac{\d N^{\rm{HNL}}}{\d E}\right|^{\rm{norm=1}}$ refers  to the energy spectrum normalized to unity. An analogous expression holds for HNL production in muon decay, accounting for the fraction of muons produced via pion decay that subsequently decay at rest.

Since pion decay is isotropic, the total number of HNLs reaching the detector per unit energy is
\begin{align}
    \label{eq:HNL_det}
    \left.
    \dfrac{\d N^{\rm{HNL}}}{\d E}\right|^{\rm{det}} &= \left.\dfrac{\d N^{\rm{HNL}}}{\d E}\right|^{\rm{prod}}
    \times 
    \dfrac{A_{\rm{det}}}{4 \pi L^2}\, ,
\end{align}
where $A_{\text{det}}$ is the effective cross-sectional area of the detector (assuming a cubic geometry) and $4\pi L^2$ is the area of the sphere of radius $L$ centered on the source, with $L$ the source-detector baseline. This approximation holds for all the detectors considered here, for which L is large enough compared to the typical detector size.
The relevant parameters of all detectors considered in this work are summarized in Table~\ref{tab:detectors}. Note that, for all detectors, we assume a three-year data-taking period with 5000 effective hours of operation per year. The POT number scales as  $N^{\rm{POT}} = 6.24 \times 10^{15}\left(\frac{P~[\rm MW]}{E~[\rm GeV]}\right) \Delta t[\rm s]$, $\Delta t$ being the data-taking time. This assumption corresponds to a total exposure of $N^{\rm{POT}} = 5.184 \times 10^{23}$~\cite{Chatterjee:2022mmu}. Regarding the average number of $\pi^+$ produced per POT, we assume $c_{\pi^+} = 0.11$ for all detectors, already accounting for an increase in the SNS beam power up to $\sim 2$ MW, as foreseen~\cite{COHERENT:2021yvp,COHERENT:2026ewu}.

The total number of detected decay events is then given by
\begin{equation}
    \label{eq:Ndecay}
   N_\mathrm{decay} = \int \d E  \left.
    \dfrac{\d N^\mathrm{HNL}}{\d E}\right|^\mathrm{det} \mathcal{P}_\mathrm{decay}(E) \, \mathcal{B}(N \to e^+  e^- \nu)\, \varepsilon(E) \, ,
\end{equation}
where $\varepsilon(E)$ is the detector efficiency, which we assume to be flat in energy for the present analysis\footnote{This is a simplification, as the efficiency in principle depends on the detector technology and would require a dedicated simulation of the $e^+e^-$ final state.} and $\mathcal{P}_\mathrm{decay}$ is the probability of the HNL decaying inside the detector volume,
\begin{equation}
    \mathcal{P}_\mathrm{decay}(E) = e^{-L/\lambda} \left(1 - e^{-s/\lambda}  \right) \, ,
\end{equation}
with $\lambda = \gamma(E) \beta(E) c \tau$ the laboratory-frame decay length, $\tau$ the HNL proper lifetime, and $s$ the path length traversed by the HNL inside the detector, which we assume equal to the thickness of the detector, given the assumption $s \ll L$~\cite{AristizabalSierra:2026jgp}.

The upper limit on the number of signal events, $S_{\rm up}$, at a given confidence level is obtained using a Bayesian approach for Poisson-distributed counting data. The posterior probability density for $S$ is proportional to the product of the Poisson likelihood and a prior on $S$. Given the small number of signal events expected, the likelihood is
\begin{align}
    P(N|S)=\frac{(S+B)^{N}}{N!}e^{-(S+B)},
    \label{eq:poisson}
\end{align}
where $N$ is the total number of observed events, and $S$ and $B$ are the number of signal and background events, respectively. We adopt a conservative, flat prior restricted to non-negative signal yields,,
\begin{align}
    \pi(S) = \left\{ \begin{array}{lr} 1,~S \geq 0\\ 0,~ S < 0 \end{array} \right. \, ,
\end{align}
corresponding to a Heaviside step function. The upper limit $S_{\rm up}$ on the signal events at a credibility level $1-\alpha$ is then obtained by solving~\cite{ParticleDataGroup:2026aaa}
\begin{align}
    1-\alpha=\frac{\int_{-\infty}^{S_{\rm up}}P(N|S)\pi(S)\d S}{\int_{-\infty}^{\infty}P(N|S)\pi(S)\d S}\, ,
\end{align}
which, using the incomplete gamma function, can be written as
\begin{align}
\label{eq:Sup}
    \alpha=\frac{\Gamma(N+1,S_{\rm up}+B)}{\Gamma(N+1,B)}\, .
\end{align}
We solve this expression numerically for $S_{\rm up}$ at the $1-\alpha = 90\%$ confidence level (CL). For projected sensitivities, we adopt the Asimov dataset, setting the observed number of events equal to the expected background, $N = B$, and solve Eq.~\eqref{eq:Sup} to obtain the expected upper limit $S_{\rm up}$.

A few comments are in order at this point. First, while our assumption of a flat efficiency may appear overly simplistic, in what follows we perform all sensitivity studies for both $\varepsilon = 50\%$ and $\varepsilon = 100\%$, in order to illustrate how our results vary with different assumptions. A more careful estimation of the efficiency, tailored to each specific detector configuration, is left for future work. Second, the background estimate is likewise dependent on the experimental configuration; we discuss the relevant background sources in the next subsection. Here, we simply note that we perform our sensitivity studies assuming three benchmark values: a zero-background scenario, $B = 100$, and $B = 10^5$, representing a more realistic and a very conservative background assumption, respectively.

\subsection{Signal and Background Considerations}

The signal consists of an electromagnetic (EM) energy deposition from the $e^+e^-$ pair produced in $N \to \nu e^+ e^-$ decay within the fiducial volume of the detector, in time coincidence with the SNS beam spill. The total visible energy is $T_{ee} = T_{e^+} + T_{e^-} \lesssim m_N$, distributed in the range of a few to tens of MeV depending on the HNL mass. We therefore identify three main background categories.
 
\paragraph{Beam-correlated backgrounds.}
Photons and neutrons produced at the spallation target and surviving the shielding can reach the detector volume. Energetic photons that convert to $e^+e^-$ pairs are the most concerning, as they directly mimic the signal topology in terms of EM energy deposition; however, their flux is suppressed exponentially by the $\mathcal{O}(20\,\text{m})$ of steel and concrete shielding between the target and detector. Consequently, the photon backgrounds are subdominant compared to neutrons~\cite{Cherkashyna:2014}. Fast neutrons arriving in the prompt window are suppressed by pulse-shape discrimination and could be distinguished from the EM nature of the signal, since neutron-induced nuclear recoils have a distinct response in most detector technologies.

\paragraph{Neutrino-induced backgrounds.} 
SM neutrino interactions constitute the most irreducible background. \cevns~produces nuclear recoils that are well separated from EM energy depositions in detectors with sufficient discrimination capability. Neutrino-electron elastic scattering (\eves), $\nu + e^- \to \nu + e^-$, produces a single electron in the same MeV-scale energy range as the signal; although its rate is subdominant to \cevns, it lacks an accompanying nuclear recoil and must be considered explicitly, particularly in detectors without directional or topological reconstruction. Charged-current interactions of $\nu_e$ on detector nuclei can also produce electrons in the relevant energy range, and neutral-current interactions producing $\pi^0 \to \gamma\gamma$ or nuclear de-excitation photons may contribute at sub-leading level.

\paragraph{Cosmic-ray backgrounds.} For $m_N \sim \mathcal{O}(100)$ MeV, where the highest sensitivity is reached, the visible energy of the $e^+e^-$  pair spans tens to $\sim 150\,\text{MeV}$, a range that is unfortunately populated by cosmic-ray induced processes. The beam timing coincidence provides a duty-cycle suppression of $f_{\rm duty}\sim 5\times10^{-6}$~\cite{COHERENT:2026ewu}, but several cosmic-ray induced processes deposit energy in exactly this window and must be evaluated carefully. Michel electrons from muons stopping in or near the active volume, have an endpoint ($\sim53\,\text{MeV}$) that also falls inside the expected signal region.

A precise prediction of these backgrounds would require a full Monte Carlo simulation~\cite{GEANT4:2002zbu} of the cosmic-ray flux propagating through Neutrino Alley at the SNS, including the muon veto and lead shielding, and should be calibrated using the anticoincidence data, as is standard practice for COHERENT's steady-state background characterization. Additional detector- and shielding-specific considerations may further reduce these backgrounds depending on the technology and its ability to (partially) reconstruct the topology of the $e^+e^-$ pair. While such an analysis lies beyond the scope of our phenomenological study, we obtain a conservative order-of-magnitude estimate as follows. 

We adopt a representative flux of cosmic-ray-induced photons above $\sim$ tens of MeV energies reaching the detector volume, $\Phi_\gamma \simeq 10^{-2}\,\text{cm}^{-2}\text{s}^{-1}$, chosen to conservatively include the uncertainty in the local shielding configuration and the spread in energy range, noticing that the flux is expected to rapidly drop above $\sim 50$ MeV. This value is of the same order as the total sea-level cosmic-ray flux reported in
Ref.~\cite{Bogdanova:2006ex} ($\sim3\times10^{-2}\,\text{cm}^{-2}\text{s}^{-1}$, dominated by the muon component). We adopt it as a deliberately conservative ceiling for the photon flux specifically, without applying either the suppression expected from the 8~meter-water-equivalent overburden or the reduction associated with isolating the photon component from the total flux, so as to remain conservative. We then note that all detectors under study are several radiation lengths thick along their characteristic dimension, so that essentially any $\gtrsim$ MeV photon entering the active volume converts to an $e^+e^-$ pair ($P_{\rm conv}\gtrsim98\%$) and therefore mimics the signal topology at the level of total
EM energy deposition. The expected count is then
\begin{equation}
B \;\simeq\; \Phi_\gamma \times A_{\rm det}\times P_{\rm conv}\times \Delta t \times f_{\rm duty},
\label{eq:cosmic_bkg}
\end{equation}
where $A_{\rm det}$ is the effective cross-section for an isotropic flux incident on a detector of characteristic size $L$, $\Delta t$ is the exposure time, and $f_{\rm duty}\sim 5\times10^{-6}$ is the SNS beam-spill duty cycle~\cite{COHERENT:2026ewu}, which strongly suppresses this steady-state background relative to the beam-correlated signal window. Unlike beam-related or intrinsic radioactive backgrounds, this cosmic-ray background tracks the detector's geometric size rather than its mass, favoring compact detector designs; it also motivates the muon veto and lead shielding described above as the primary handles for suppressing this background in situ. Evaluating Eq.~\eqref{eq:cosmic_bkg} for the detectors in Table~\ref{tab:detectors} over $\Delta t=15{,}000$~hr, yields $B\sim [10^4,10^5]$. spanning the most compact detectors to the largest (LAr-10t).
We stress that the absolute normalization above carries a large uncertainty and the resulting range is meant as a conservative input for assessing the impact of this background on our sensitivity projections, rather than a precise prediction. In the main text we present results under the assumption of a background-free search, and in Appendix~\ref{app:moreresults} we additionally show results for two illustrative values spanning this range: $B=100$, representative\footnote{A full simulation of cosmic-ray backgrounds for the heavy-water detector at the SNS predicts $\mathcal{O}(10^2)$ residual cosmic-ray events in a compatible energy window and data-taking time, after muon-veto and cut selection~\cite{COHERENT:2021xhx}.} of a more realistic scenario once shielding
and veto suppression are taken into account, and $B=10^5$, corresponding to the most conservative end of our estimate.

\subsection{Other facilities}
\label{subsec:otherSS}

\begin{table*}[!htb]
\renewcommand{\arraystretch}{1.3}
\begin{tabular}{|l|l|c|c|c|c|c|}
\toprule
\bf{Facility} & \bf{Detector} & $N^{\mathrm{POT}}/\text{yr}$ & $c_{\pi^+}$ & Mass [kg] & Size [cm] & Baseline [m] \\
\toprule
ESS~\cite{Baxter:2019mcx} & Xe   &  &  & 20   &  50  & 20  \\ \hhline{|~|-|~|~|-|-|-|}
 & CsI   & \multirow{-2}{*}{$2.8\times10^{23}$} & \multirow{-2}{*}{0.15}  & 22.5   &  17  & 20  \\
\hline
J-PARC~\cite{Collar:2025sle} & Xe &  &  & 20 & 50   & 20 \\ \hhline{|~|-|~|~|-|-|-|}
 & CsI & \multirow{-2}{*}{$3.5 \times 10^{22}$} & \multirow{-2}{*}{0.24} & 22.44 & 21.4   & 20 \\
\hline
CSNS~\cite{Su:2023klh} & CsI & $ 9.8 \times 10^{21}$ & 0.12 & 300 & 20   & 10.5 \\
\hline
\bottomrule
\end{tabular}
\caption{Detector parameters at other spallation source facilities considered in this work: the ESS, J-PARC, and the CSNS. For each detector we list the annual number of protons on target $N^{\mathrm{POT}}/\mathrm{yr}$, the average number of $\pi^+$ produced per proton on target $c_{\pi^+}$, the detector mass, linear size, and baseline distance $L$ from the target.}
\label{tab:other_facilities}
\end{table*}

We have so far focused on the COHERENT experimental program at the SNS to explore the expected sensitivity to HNL searches. However, other spallation source facilities are currently in operation or under development worldwide. In this section, we briefly discuss the potential of these alternative installations for HNL searches and compare them to those presented for the SNS.

We begin with the European Spallation Source (ESS), currently under construction in Lund, Sweden, that aims to become the most intense proton beam source for multi-disciplinary research~\cite{Abele:2022iml}. As a spallation source, the dynamics of particle production at the ESS is similar to that at the SNS, with a few differences relevant for a complementary HNL search.
In particular, the ESS will employ a tungsten target with a 14~Hz proton beam pulse at 2~GeV and an expected beam power of 5~MW~\cite{Abele:2022iml}. Under these conditions, a $c_{\pi^+}$ of 0.15 can be reached, substantially enhancing pion production with respect to the current SNS configuration. An initial proposal to measure \cevns~at the ESS considers six different detector technologies, of which we identify Xe and CsI targets~\cite{Baxter:2019mcx, Monrabal:M7s} as the most promising  ones for HNL searches, given their large target masses and well-established response to MeV-scale electromagnetic energy depositions; the remaining proposed technologies are either optimized for low-threshold \cevns\ measurements or feature significantly smaller fiducial masses, reducing their sensitivity to the $e^+e^-$ signal of interest here. We therefore compute the expected HNL sensitivities for these three  detectors, assumed to be located at a baseline of 20 m, with masses and sizes as listed in Table~\ref{tab:other_facilities}.

Another relevant facility is the China Spallation Neutron Source (CSNS), currently in operation in Dongguan, China~\cite{Su:2023klh}. The CSNS likewise employs a tungsten target, with a pulse rate of 25~Hz, a proton beam energy of 1.6~GeV, and an initial beam power of 0.14~MW that is planned to be upgraded to 0.5~MW. 
A proposal to measure \cevns~with a CsI detector is currently under development, with data taking expected to begin soon~\cite{CICENNS2025talk}. Given the current beam characteristics, the production of charged pions corresponds to $c_{\pi^+} = 0.12$; the proposed detector mass is also listed in Table~\ref{tab:other_facilities}. 

Finally, as a third example we consider the Japan Proton Accelerator Research Complex (J-PARC)~\cite{JSNS2:2013jdh}. For this experimental setup, a mercury target is used with a 3~GeV proton beam, a power supply of 1~MW, and a pulse structure of 25~Hz double pulses with a $0.1\,\mu\mathrm{s}$ separation, resulting in $c_{\pi^+} = 0.24$. The proposed \cevns~program at J-PARC also considers Xe and CsI detection technologies~\cite{Collar:2025sle}, with expected baselines of 20 m, and expected detector masses as listed in Table~\ref{tab:other_facilities}. 

We study the projected HNL sensitivities across these facilities and corresponding proposed detectors, for all of them assuming a 3-year data-taking time, with 5000 effective hours of operation per year.

\section{Results}
\label{sec:results}

\begin{figure*}[htb!]
    \centering
    \includegraphics[scale=0.45]{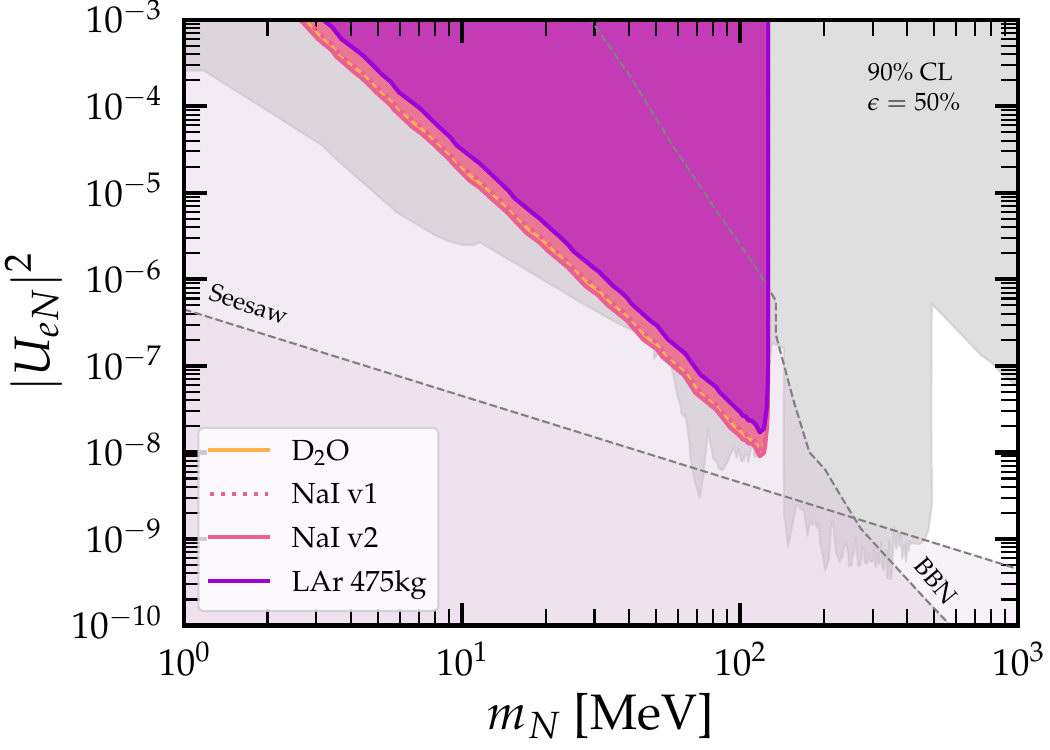}
    \includegraphics[scale=0.45]{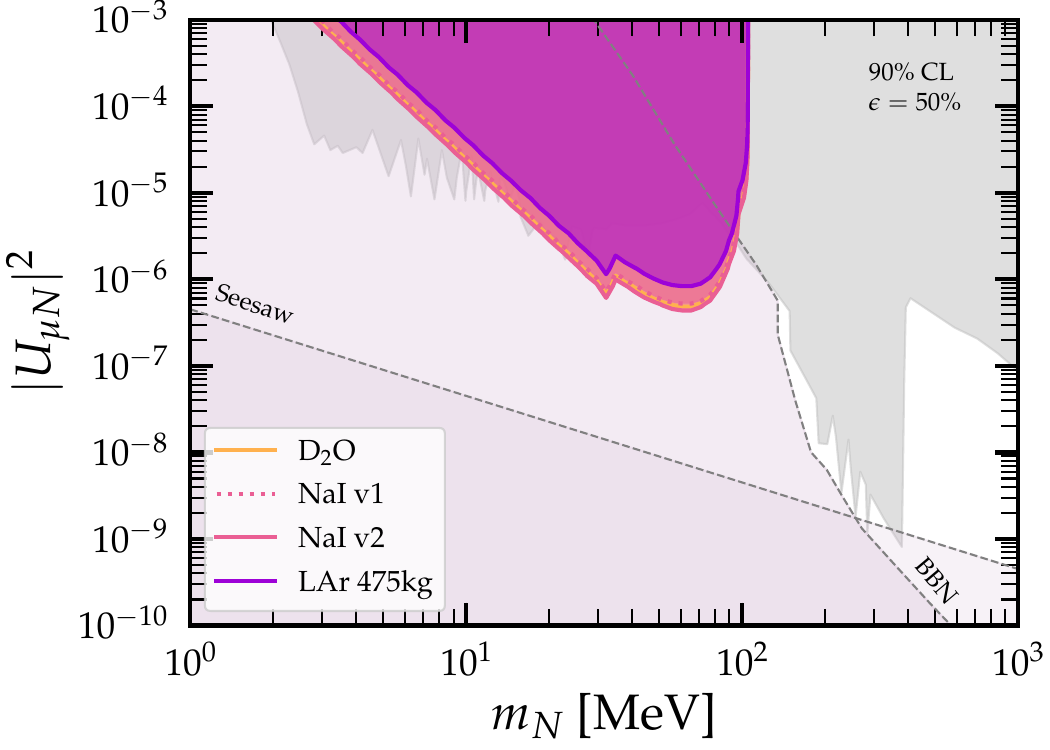}\\
    \includegraphics[scale=0.45]{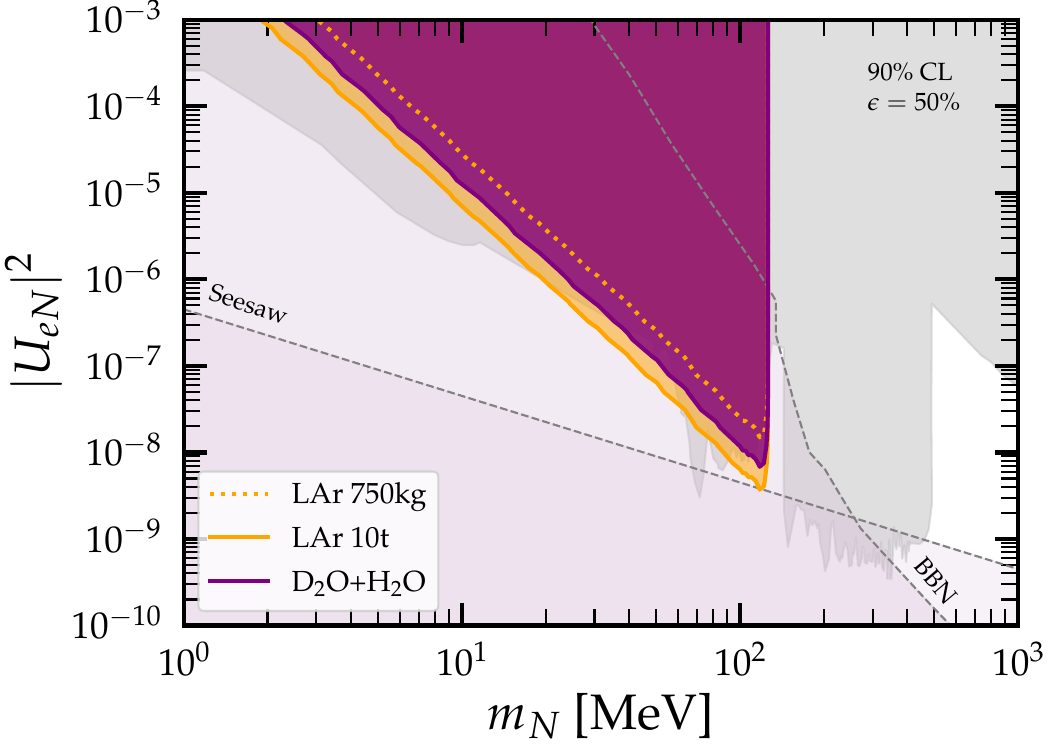}
    \includegraphics[scale=0.45]{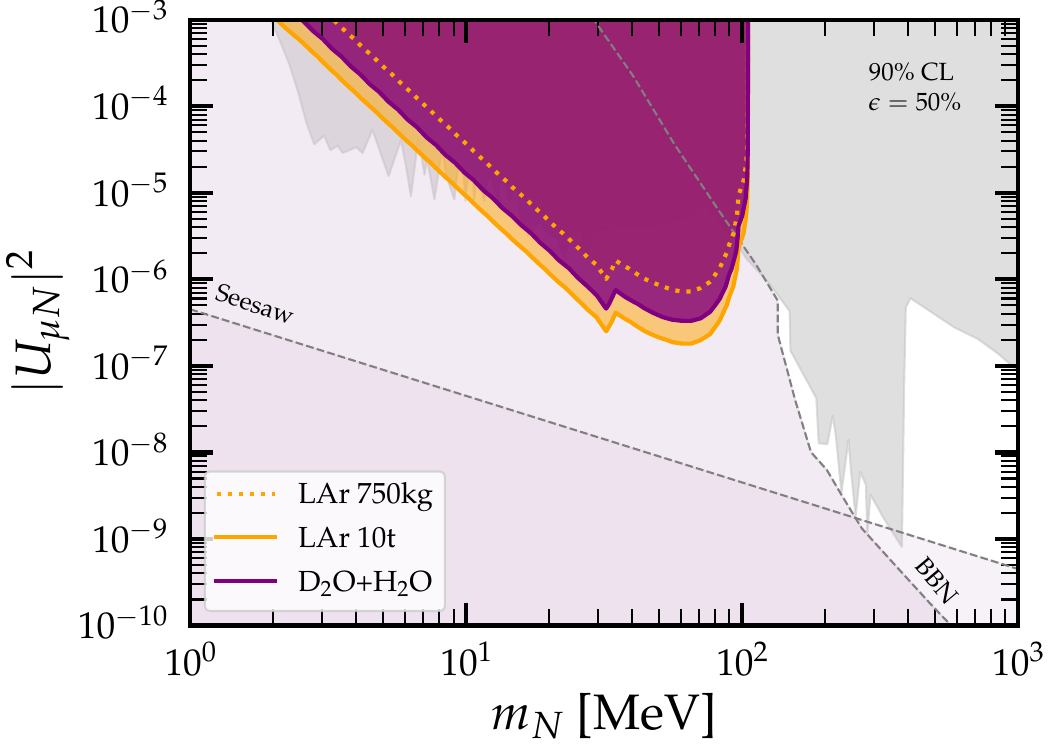}
    \caption{Projected 90\% CL exclusion sensitivity of currently operating or soon-to-be-deployed (\textbf{upper panels}) and future (\textbf{lower panels}) COHERENT detectors in the $(m_N, |U_{\ell N}|^2)$ plane, for the electron-mixing (\textbf{left panel}) and muon-mixing (\textbf{right panel}) benchmark scenarios. See Tab.~\ref{tab:detectors} for more details on the SNS detectors. We assume a flat detector efficiency of $50\%$ and the background-free hypothesis (corresponding to $S_{\rm up} = 2.3$). The gray shaded region shows existing laboratory constraints, as discussed in the main text. The light gray dashed lines indicate cosmological (BBN) bounds and the target mixing values predicted by the canonical seesaw mechanism. See the main text for further details.}
    \label{fig:SNS_50eff}
\end{figure*}

In this section, we discuss the inferred sensitivities to HNLs with masses in the MeV-GeV range. Figure \ref{fig:SNS_50eff} shows the 90\% CL projected exclusion regions, in the $(m_N, |U_{\ell N}|^2)$ (for $\ell=e,\mu$) parameter space for the COHERENT detectors at the SNS listed in Table~\ref{tab:detectors}. For clarity, we split the detectors in two categories according to their operational status: those currently running or expected to begin taking data soon (NaI v1 and v2, D$_2$O, and LAr-475\,kg, upper panels), and those planned as future detectors (D$_2$O+H$_2$O, LAr-750, and LAr-10t, lower panels). Throughout this figure, we assume a flat efficiency $\epsilon=50\%$, and the most-optimistic, background-free hypothesis. We provide in Appendix~\ref{app:moreresults} the same projected sensitivities, but with different efficiency and background assumptions.

As benchmark scenarios, we consider in Fig.~\ref{fig:SNS_50eff} two mixing cases: $i)$ only electron mixing is activated (left panels), and $ii)$ only muon mixing is activated (right panels). These simplified single-flavor benchmarks, 
even though inconsistent~\cite{Drewes:2022akb} with the observed properties of light active neutrinos~\cite{deSalas:2020pgw,Esteban:2024eli,Capozzi:2025wyn}, are extensively used in phenomenological studies and allows us to easily compare with other searches. The overall shape of the exclusion contour for a given mixing scenario is common to all detectors, since the relevant production and decay channels are the same in each case.
In the electron-mixing scenario, the sensitivity is dominated by pion decay in the whole mass range considered and terminates near the pion mass ($m_N \lesssim m_{\pi} \approx 140$~MeV), since pion decay at rest cannot kinematically produce heavier HNLs. The muon-mixing scenario instead exhibits two distinct regimes: at low $m_N$, the sensitivity is again dominated by pion decay, while above $m_N \approx 34$~MeV, where the two-body channel $\pi^+ \to \mu^+ N$ becomes kinematically forbidden, it transitions to being dominated by the three-body muon decay channel. Correspondingly, the muon-mixing sensitivity terminates near the muon mass ($m_N \lesssim m_{\mu} \approx 105$~MeV).

In the same figures, we also show the most relevant existing laboratory limits~\cite{Fernandez-Martinez:2023phj}, depicted as gray shaded areas. For the electron-mixing case, the most constraining ones come from $^{20}$F $\beta$-decay~\cite{Bryman:2019bjg}, super-allowed $\beta$-decay~\cite{Bryman:2019bjg}, pion universality~\cite{Bryman:2019bjg}, Borexino~\cite{Borexino:2013bot}, kaon universality, LSND~\cite{Hostert:2025ffy} (see also~\cite{Fernandez-Martinez:2023phj}), TRIUMF~\cite{Britton:1992xv}, PIENU~\cite{PIENU:2017wbj}, NA62~\cite{NA62:2020mcv}, T2K~\cite{T2K:2019jwa}, Belle, BEBC~\cite{WA66:1985mfx,Barouki:2022bkt}, and CHARM~\cite{CHARM:1985nku}; for the muon-mixing case instead, they come from PSI~\cite{Daum:1987bg}, PIENU~\cite{PIENU:2019usb}, MicroBooNE~\cite{Kelly:2021xbv}, BNL-E949~\cite{BNL-E949:2009dza}, T2K~\cite{Bernardi:1985ny,Bernardi:1987ek,Arguelles:2021dqn}, NA62~\cite{NA62:2021bji}, KEK~\cite{Hayano:1982wu,Yamazaki:1984sj}, BEBC~\cite{WA66:1985mfx}, CHARM~\cite{CHARM:1985nku}, and NuTeV~\cite{NuTeV:1999kej}.

\begin{figure*}[!ht]
    \centering
    \includegraphics[scale=0.45]{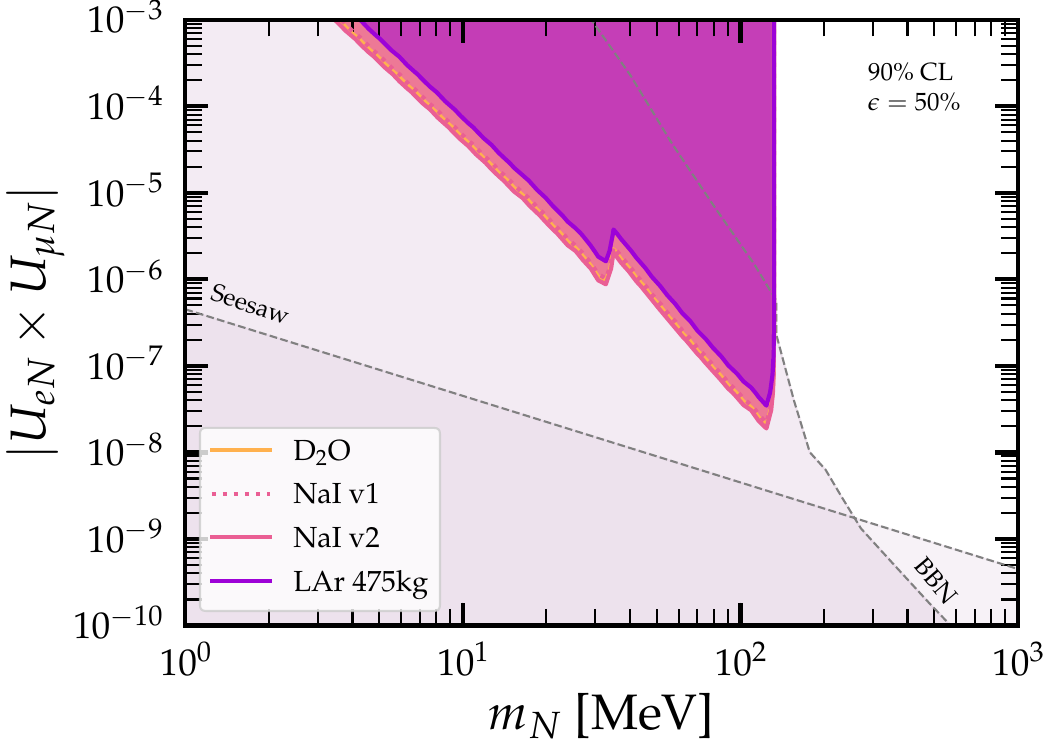}
    \includegraphics[scale=0.45]{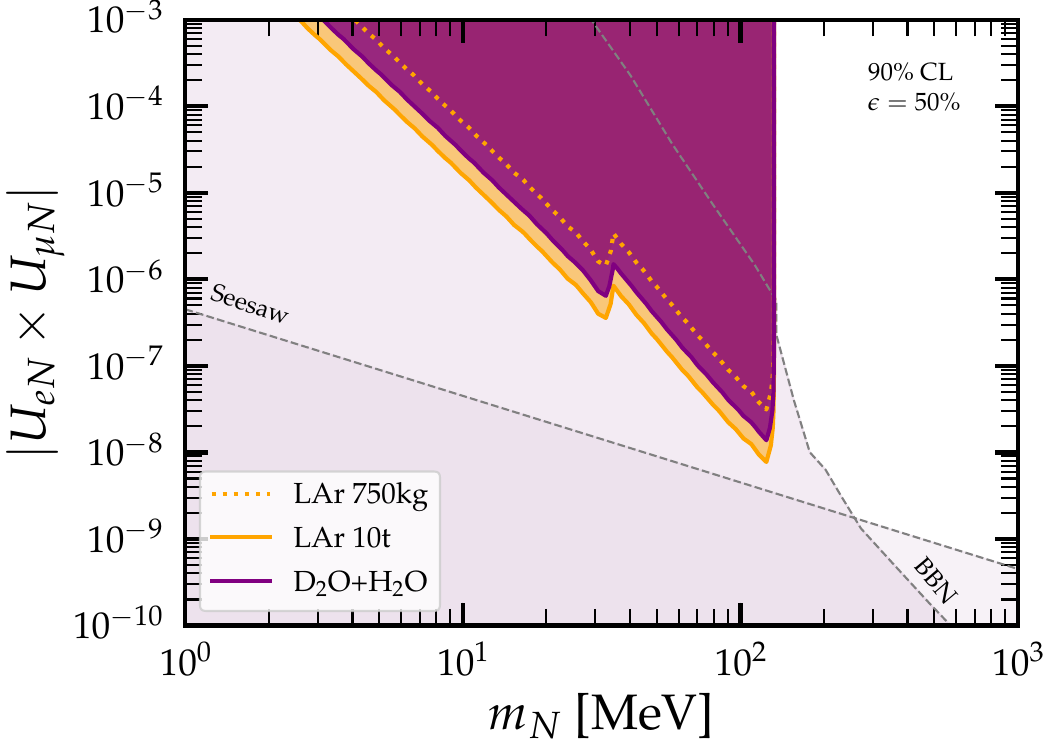}
    \caption{Projected 90\% CL exclusion sensitivity in the $(m_N, |U_{eN} \times U_{\mu N}|)$ plane, assuming the benchmark scenario with equal mixing to electrons and muons, for current (\textbf{left panel}) and future (\textbf{right panel}) COHERENT detectors. We assume a flat detector efficiency of $50\%$ and the background-free hypothesis (corresponding to $S_{\rm up} = 2.3$).
    }
    \label{fig:mixed_50eff}
\end{figure*}

\begin{figure*}[t!]
    \centering
    \includegraphics[scale=0.45]{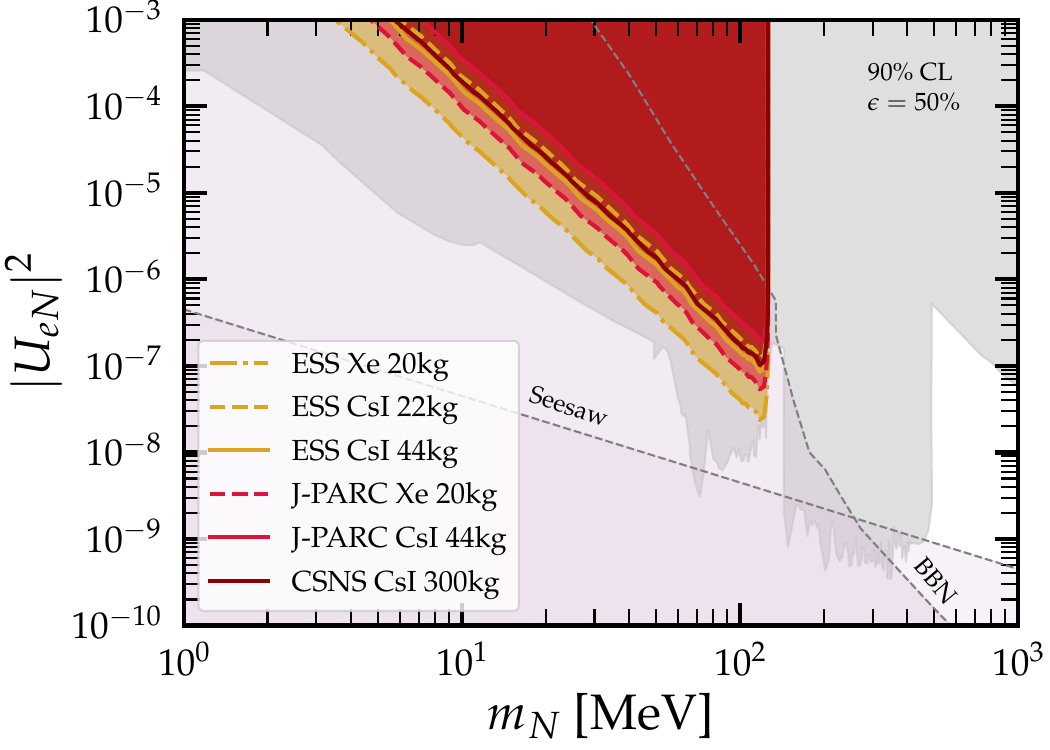}
    \includegraphics[scale=0.45]{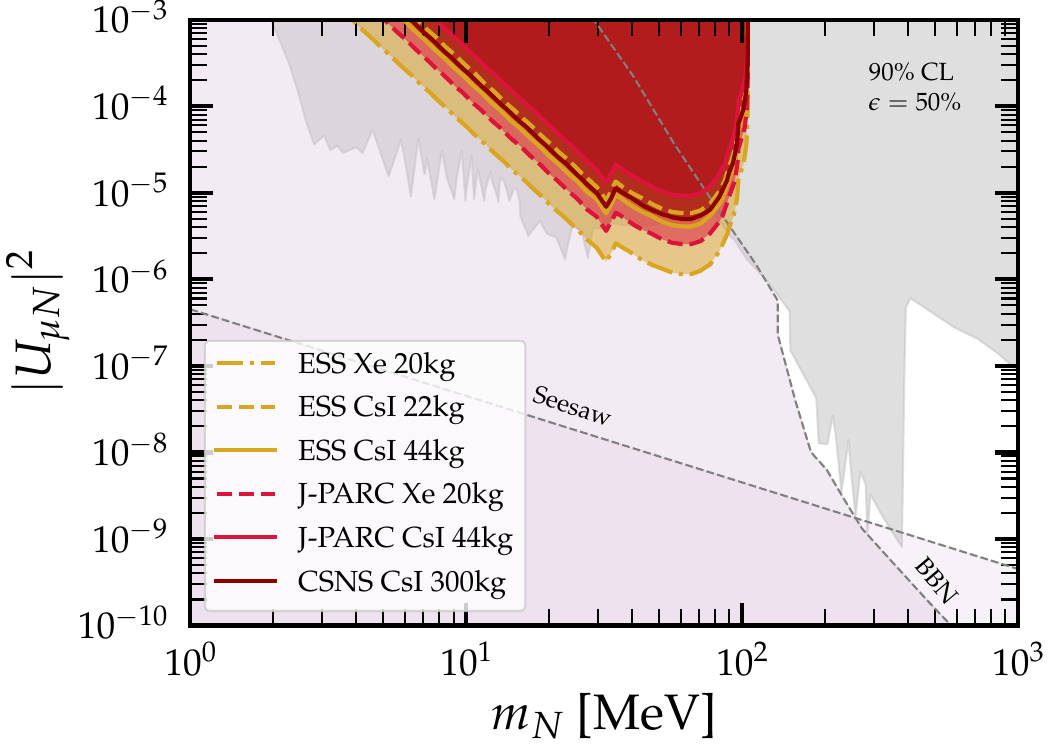}
  \caption{Projected 90\% CL exclusion sensitivity of the detectors considered at other spallation source facilities -- ESS, J-PARC, and CSNS -- in the $(m_N, |U_{\ell N}|^2)$ plane, for the electron-mixing (\textbf{left panel}) and muon-mixing (\textbf{right panel}) benchmark scenarios. See Table~\ref{tab:other_facilities} for the corresponding detector parameters. We assume a flat detector efficiency of $50\%$ and the background-free hypothesis (corresponding to $S_{\rm up} = 2.3$). The gray shaded region shows existing laboratory constraints, as discussed in the main text. The light gray dashed lines indicate cosmological (BBN) bounds and the target mixing values predicted by the canonical seesaw mechanism.} \label{fig:otherfac_50eff}
\end{figure*}

Beyond laboratory constraints, the mixing of HNLs with active neutrinos can also affect the cosmological evolution of the early Universe. 
In particular, HNLs with masses $m_N \lesssim 1$ GeV can be sufficiently long-lived to modify the abundances of light nuclei formed during Big Bang Nucleosynthesis (BBN)~\cite{Boyarsky:2009ix,Ruchayskiy:2012si}, while recombination-era observables such as the Cosmic Microwave Background (CMB) and Baryon Acoustic Oscillations (BAO) provide further constraints on the HNL mixing~\cite{Vincent:2014rja}.
Although such cosmological bounds can provide the most stringent constraints on $|U_{\ell N}|^2$ for $m_N$ below the GeV scale~\cite{Kusenko:2009up,Abdullahi:2022jlv,Bolton:2019pcu}, they rely on the assumption that the standard cosmological model holds at the relevant early-Universe epochs, and are therefore subject to different, partly model-dependent assumptions compared to laboratory probes. We nonetheless show them in our Figures \ref{fig:SNS_50eff} -- \ref{fig:otherfac_50eff} for indicative purposes, as a light gray line with a light purple shaded region.
For completeness, we also indicate with a diagonal dashed line the values in the $(m_N, |U_{\ell N}|^2)$ plane for which the canonical seesaw relation $|U_{\ell N}|^2 = m_\nu / m_N$ is satisfied, taking $m_\nu = 0.05$~eV as a representative active neutrino mass scale.
Turning to the results themselves, we note that the current SNS detectors -- D$_2$O, NaI v1, LAr (476~kg), and NaI v2 -- achieve sensitivities near the kinematic edge $m_N \simeq m_\pi$ comparable to those of PIENU and NA62 for the electron mixing, whereas they allow to probe an unconstrained region of parameter space in the case of muon mixing, allowing to improve upon current bounds from T2K and MicroBooNE. Future detectors, by contrast, will be able to improve the sensitivity reach considerably, leveraging their larger volumes to probe previously unexplored regions of parameter space in the range $m_N \simeq 30$--$100$~MeV, in both mixing scenarios and particularly in the muon-mixing case for the LAr-based detectors. Sensitivities with larger efficiency  and a more conservative estimate of backgrounds are given in Appendix~\ref{app:moreresults}. The results hold qualitatively under these more realistic background assumptions, although in the most conservative case ($B=10^5$) the projected sensitivities are, in most cases, limited to regions already excluded by existing bounds.

We now compare our results with previous studies of HNL sensitivities from $\pi$-DAR in the literature. Our results are comparable to those obtained for PIP2-BD~\cite{Ema:2023buz}: in the electron-mixing case we find improved sensitivity for $m_N \lesssim 100$\,MeV, while in the 
muon-mixing case the two analyses agree closely. Compared with the COHERENT study of Ref.~\cite{Hostert:2025ffy}, our work differs in two respects: we perform a more detailed analysis of individual detector capabilities rather than a global treatment of the full detector suite, and we include HNL production via pion decays in all benchmarks, which enhances 
the sensitivity particularly in the electron-mixing scenario.
Finally, the PROSPECT collaboration has recently presented sensitivity projections for HNLs using a ton-scale hydrocarbon scintillator detector~\cite{PROSPECT:2026jsl}. The reach of currently operating or soon-to-be-deployed COHERENT detectors, 
shown in Fig.~\ref{fig:SNS_50eff}, is comparable to the PROSPECT projections, while future upgrades explored in both studies cover a larger and complementary portion of the parameter space. We however note that the PROSPECT analysis adopts different background scenarios and includes a dedicated study of the detector 
response, making a direct quantitative comparison non-trivial.

We next consider a benchmark scenario more closely motivated~\cite{Drewes:2022akb} by neutrino oscillation data, symmetry arguments, and cosmological implications: the case in which both electron and muon mixings are simultaneously activated and equal, i.e., \ $U_{eN} = U_{\mu N}$. We show in Fig.~\ref{fig:mixed_50eff} the projected sensitivities at the SNS detectors for this scenario, in the $(m_N, |U_{eN} \times U_{\mu N}|)$ plane. In this case, the sensitivity is dominated by pion decay throughout the full mass range, and terminates near $m_N \simeq m_\pi$ for the same kinematic reasons as in the single-mixing cases. A slight feature appears around $m_N \approx 34$~MeV, corresponding to the threshold at which the channel $\pi^+ \to \mu^+ N$ becomes kinematically forbidden, marking the transition between the two pion-decay production regimes.

We then turn to the sensitivity expected at other spallation source facilities. As anticipated in Sec.~\ref{subsec:otherSS}, analogous searches can be carried out at \cevns~detectors planned for deployment at the ESS, J-PARC, and the CSNS. Figure~\ref{fig:otherfac_50eff} shows the 90\% CL projected exclusion regions in the $(m_N, |U_{\ell N}|^2)$ plane, for $\ell = e, \mu$, for all detectors listed in Table~\ref{tab:other_facilities}. Also throughout this figure, we assume a flat detector efficiency of $\varepsilon = 50\%$ and the background-free hypothesis. The shape and endpoints of the exclusion regions follow the same considerations as for the SNS detectors, since both the production and detection channels are identical. We likewise overlay existing laboratory bounds (gray shaded regions) and cosmological bounds (purple shaded regions) for comparison. We find that, while the currently planned detectors at these facilities could probe interesting regions of parameter space in the electron-mixing case, they are generally not able to improve upon existing limits. In the muon-mixing case, instead, the Xe and CsI detectors at the ESS, J-PARC, and CSNS could yield competitive sensitivities in the mass range $m_N \simeq 50$--$100$~MeV.

More generally, we expect a better sensitivity reach in the muon-mixing scenario due to the larger muon-neutrino component in the flux from $\pi$-DAR. We also note that these facilities do not provide sensitivity to tau-flavor mixing, given the beam energies involved. The goal of the present work has been to provide simplified yet realistic predictions for testing HNL decays at \cevns-based detectors using spallation-source neutrinos. Looking ahead, several experimental handles can improve the sensitivity of these searches: $i)$ larger values of $N^{\rm POT}$, achievable through higher beam power or longer running times, directly translate into larger neutrino and consequently HNL fluxes; $ii)$ larger detector volumes increase the available in-detector decay path, making ton-scale detectors particularly well suited, especially when combined with short source-detector baselines that mitigate the geometric $1/(4\pi L^2)$ suppression; $iii)$ experimental improvements such as higher detection efficiency, dedicated background reduction strategies exploiting the timing structure of the beam, shielding, and other detector-specific techniques will further extend the reach of these by-product HNL searches at facilities primarily designed for \cevns~measurements.

\section{Conclusions}
\label{sec:conclusions}

The recent experimental efforts aimed at measuring the \cevns~process with increasing precision and different target materials have opened new opportunities for compact neutrino detectors exploiting low-energy, high-intensity neutrino fluxes. In particular, spallation-source experiments such as those of the COHERENT Collaboration have demonstrated the remarkable versatility of this class of detectors and their extended potential for new physics searches beyond their primary mission. 

In this context, we have explored the sensitivity of \cevns-based detectors to the production and decay of heavy neutral leptons (HNLs) with masses in the MeV-GeV range, accessible at spallation neutron source facilities. We have considered their production from pion and muon decay at rest, and their subsequent decay into an electron-positron pair within the detector volume. We have performed phenomenological analyses assuming different mixing scenarios, and, under both conservative and optimistic assumptions of background and efficiency, derived projected sensitivities for several COHERENT detectors, either currently operating or soon-to-be deployed, and future proposed ones. We have shown that current detectors can already set interesting bounds in the $(m_N, |U_{\ell N}|^2)$ parameter space, complementary to existing laboratory constraints and, in some cases, competitive with them, provided that a sufficient reduction of backgrounds can be achieved. Future detectors, in particular the ton-scale LAr modules, could reach considerably stronger sensitivities thanks to their much larger fiducial volumes. 

We have further explored the sensitivity of \cevns~detectors planned at other spallation source facilities, namely the ESS, J-PARC, and the CSNS. These experiments could also probe relevant regions of HNL parameter space, with the best projected reach achieved by the Xe detector planned at the ESS or at J-PARC, which could yield competitive bounds in the muon-mixing case for $m_N \simeq 50$--$100$~MeV.

Taken together, our results show that spallation-source \cevns~experiments can provide meaningful and complementary constraints on HNL decays as a natural by-product of their primary physics program. Fully exploiting this potential synergy will require more detailed studies, incorporating precise energy-dependent detector efficiencies, beam-timing information, optimized signal-identification strategies, and dedicated evaluations of backgrounds and their possible mitigation. Beyond the capabilities of existing \cevns~instruments, our results also motivate the development of new dedicated compact detectors designed to probe rare HNL decays. A detector optimized for both goals—combining fine‑grained tracking or imaging, fast timing, and low‑noise calorimetry—would substantially enhance sensitivity to BSM physics. Such an instrument could leverage the intense, pulsed neutrino flux at spallation sources to perform precision \cevns~studies while also opening a new window onto long‑lived neutral particles produced in the same beam. We leave such efforts to future work, as the relevant detectors come online, anticipating results that may either establish new limits or, optimistically, reveal, the first signs of HNLs in low-energy neutrino experiments.

\section*{Acknowledgments}
We are grateful to Pablo M. Candela for useful discussions and collaboration in the early stages of this work. We also warmly thank Salvador Urrea, Pilar Coloma, Francesc Monrabal, Christoph Ternes, and Phil Barbeau for helpful comments and discussions. A.M.G. is also grateful to the members of the theory group at the Laboratoire de Physique de Clermont Auvergne for their kind hospitality and insightful conversations during the final stages of this work.\\
V.D.R., A.M.G., and V.M.L. acknowledge financial support by the grant CIDEXG/2022/20 (from Generalitat Valenciana) and by the Spanish grants CNS2023-144124 (MCIN/AEI/10.13039/501100011033 and “Next Generation EU”/PRTR), PID2023-147306NB-I00, and CEX2023-001292-S (MCIU/AEI/10.13039/ 501100011033).

\appendix

\section{Results at the SNS with different background and efficiency assumptions}
\label{app:moreresults}

\begin{figure*}[!t]
    \centering
    \includegraphics[scale=0.45]{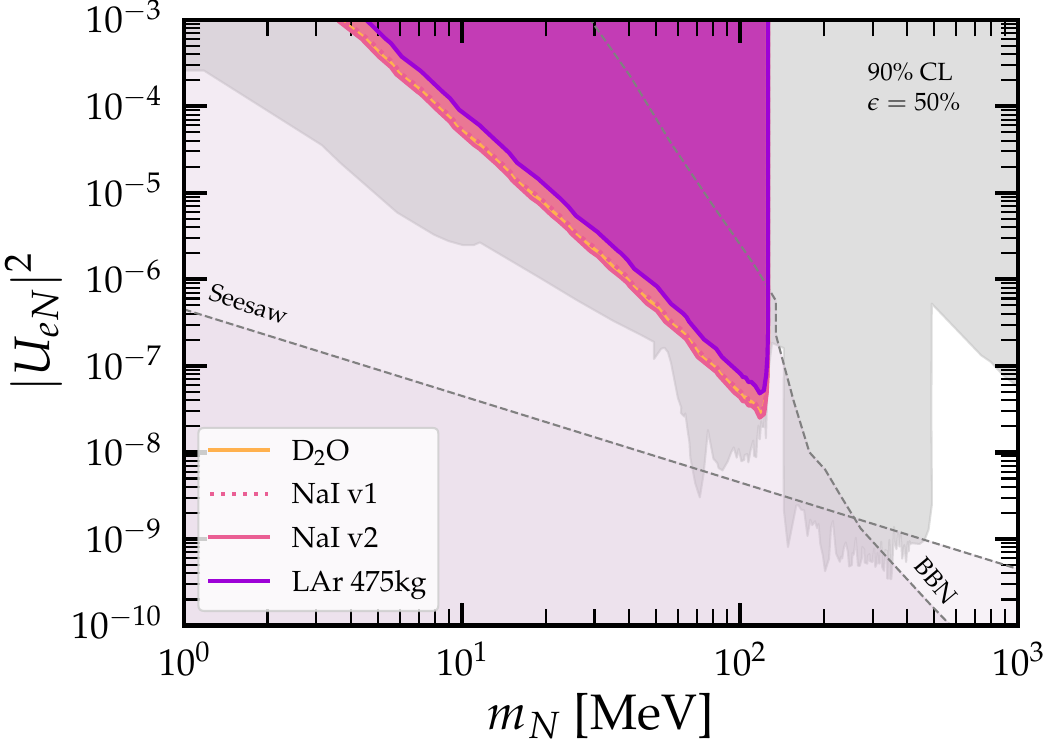}
    \includegraphics[scale=0.45]{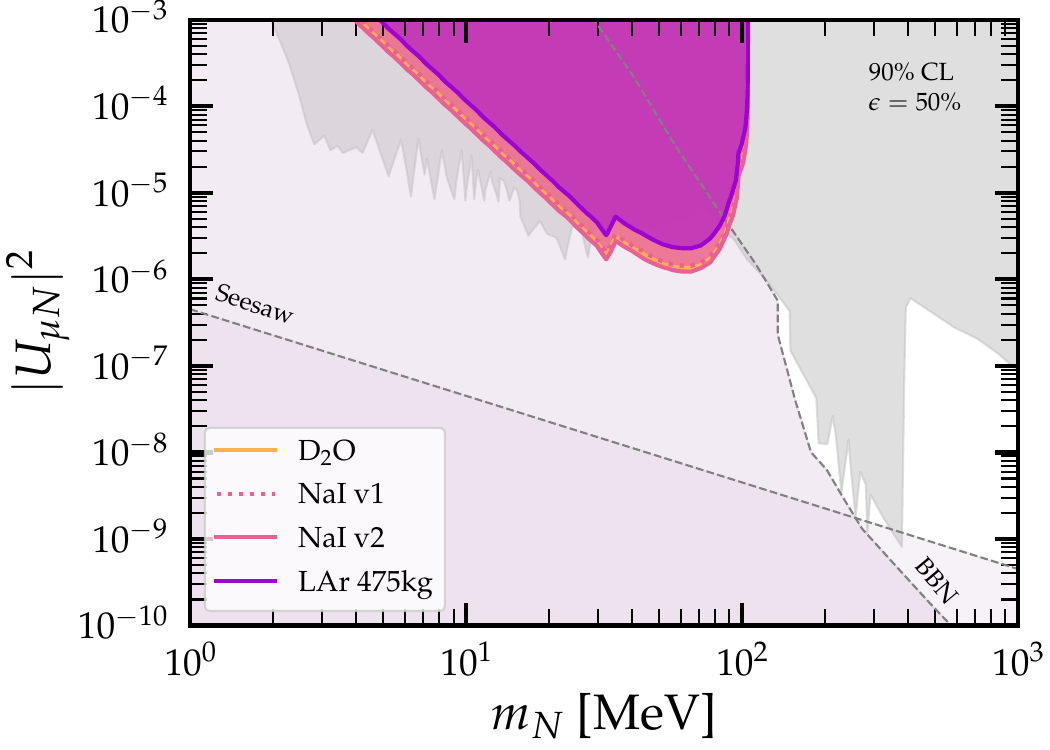}\\
    \includegraphics[scale=0.45]{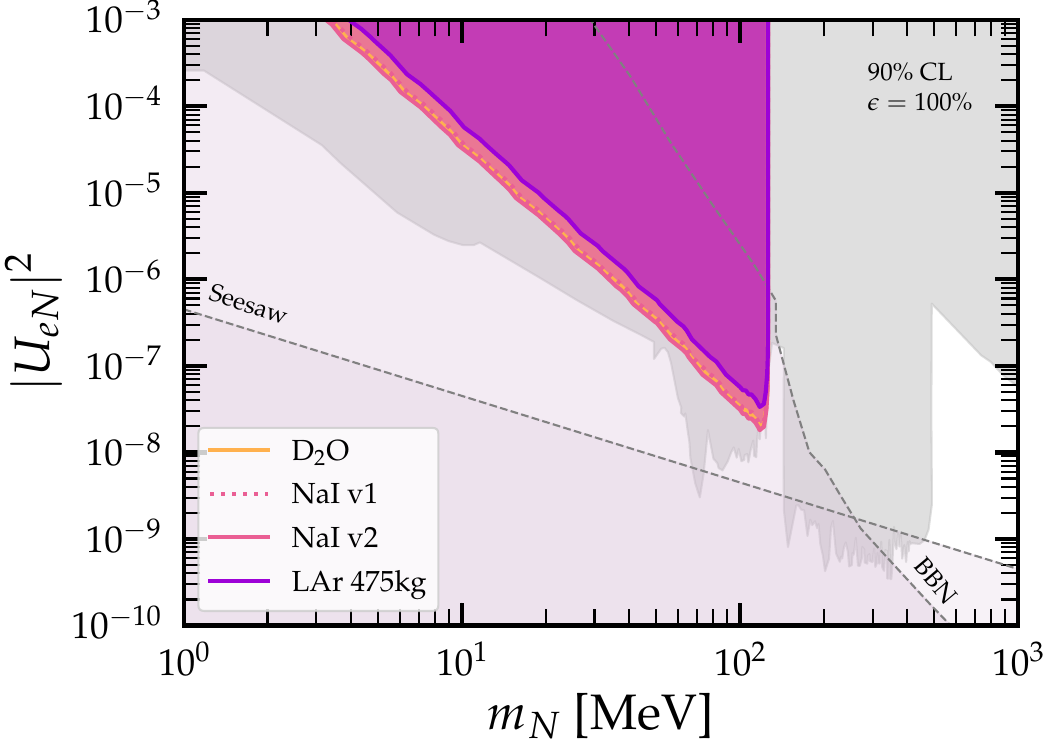}
    \includegraphics[scale=0.45]{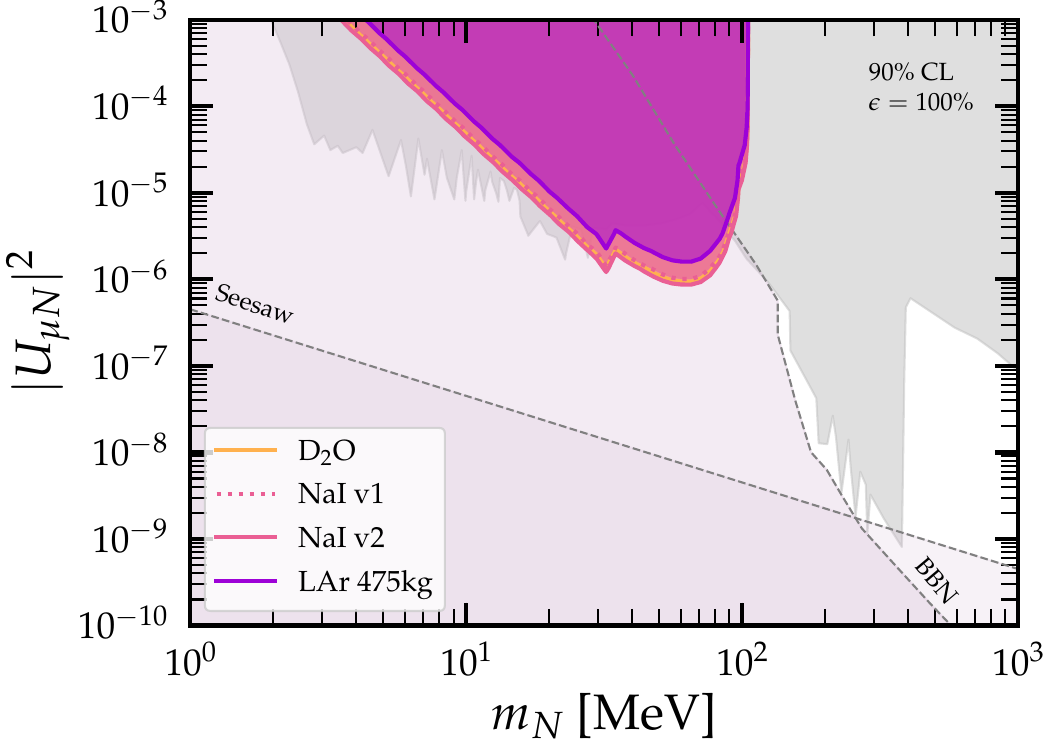}
    \caption{Same as Fig.~\ref{fig:SNS_50eff} upper panels, but assuming a total number of background events $B = 100$ (corresponding to $S_{\rm up} = 17.8$), with $\varepsilon = 50\%$  (\textbf{upper panels}) and $\varepsilon = 100\%$ efficiency (\textbf{lower panels}). }
    \label{fig:current_B100}
\end{figure*}

In this appendix, we present, for completeness, the projected sensitivities at the SNS under different assumptions on the detector efficiency and background level, complementing the results shown in the main text.

Figure~\ref{fig:current_B100} shows the 90\% CL projected exclusion regions in the $(m_N, |U_{\ell N}|^2)$ plane, for $\ell = e, \mu$, for the currently operating SNS detectors listed in Table~\ref{tab:detectors}. We compare a flat efficiency of $\varepsilon = 50\%$ (upper panels) with $\varepsilon = 100\%$ (lower panels), in both cases under a more realistic hypothesis of a total number of background events $B = 100$.

\begin{figure*}[htb!]
    \centering
    \includegraphics[scale=0.45]{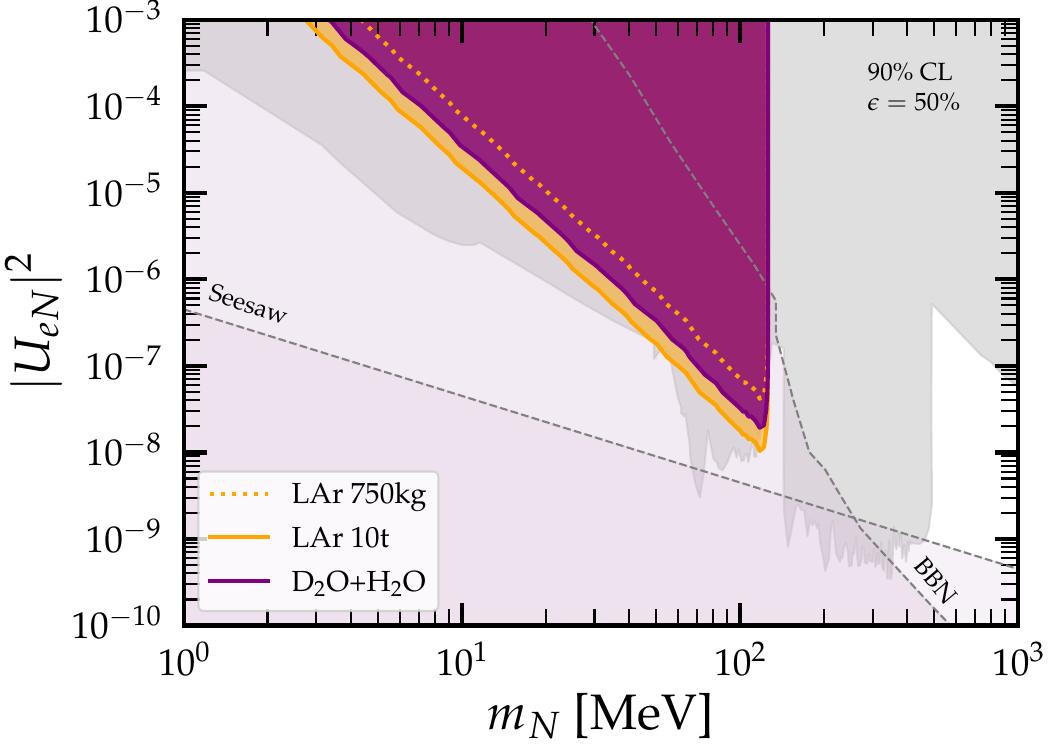}
    \includegraphics[scale=0.45]{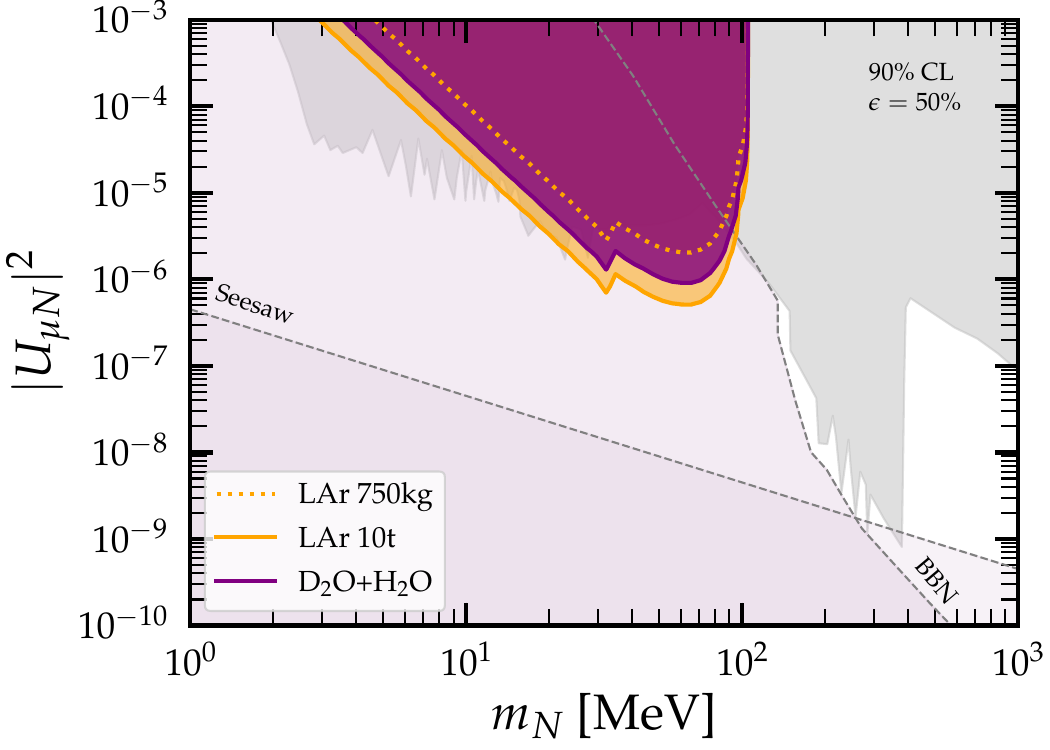}\\
    \includegraphics[scale=0.45]{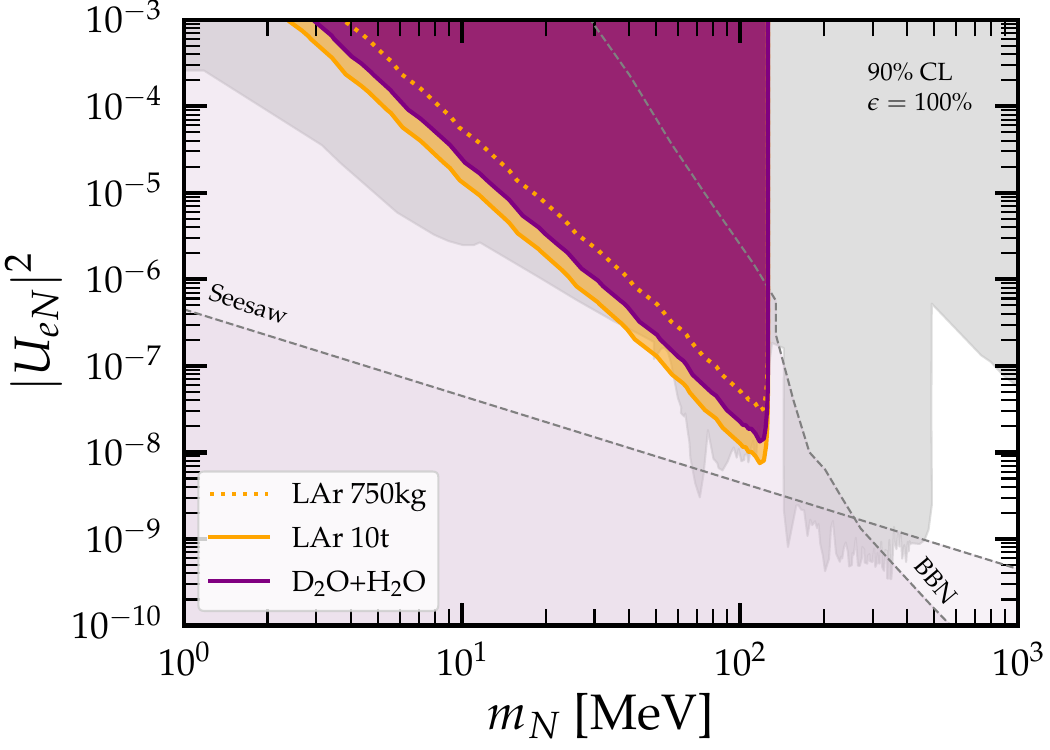}
    \includegraphics[scale=0.45]{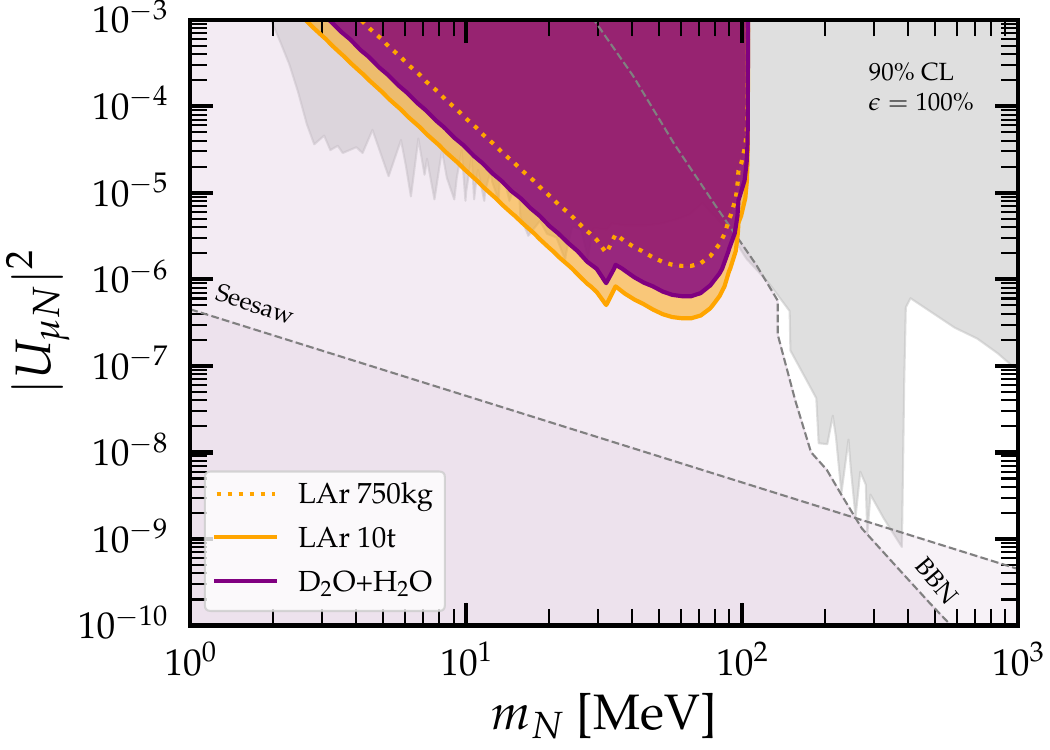}
    \caption{Same as Fig.~\ref{fig:SNS_50eff} lower panels, but assuming a total number of background events $B = 100$ (corresponding to $S_{\rm up} = 17.8$), with $\varepsilon = 50\%$  (\textbf{upper panels}) and $\varepsilon = 100\%$ efficiency (\textbf{lower panels}).}
    \label{fig:future_B100}
\end{figure*}

Figure~\ref{fig:future_B100} shows the analogous results for the future SNS detectors listed in Table~\ref{tab:detectors}, again comparing $\varepsilon = 50\%$ (upper panels) with $\varepsilon = 100\%$ (lower panels) under a more realistic hypothesis of a total number of background events $B = 100$.

\begin{figure*}[htb!]
    \centering
    \includegraphics[scale=0.45]{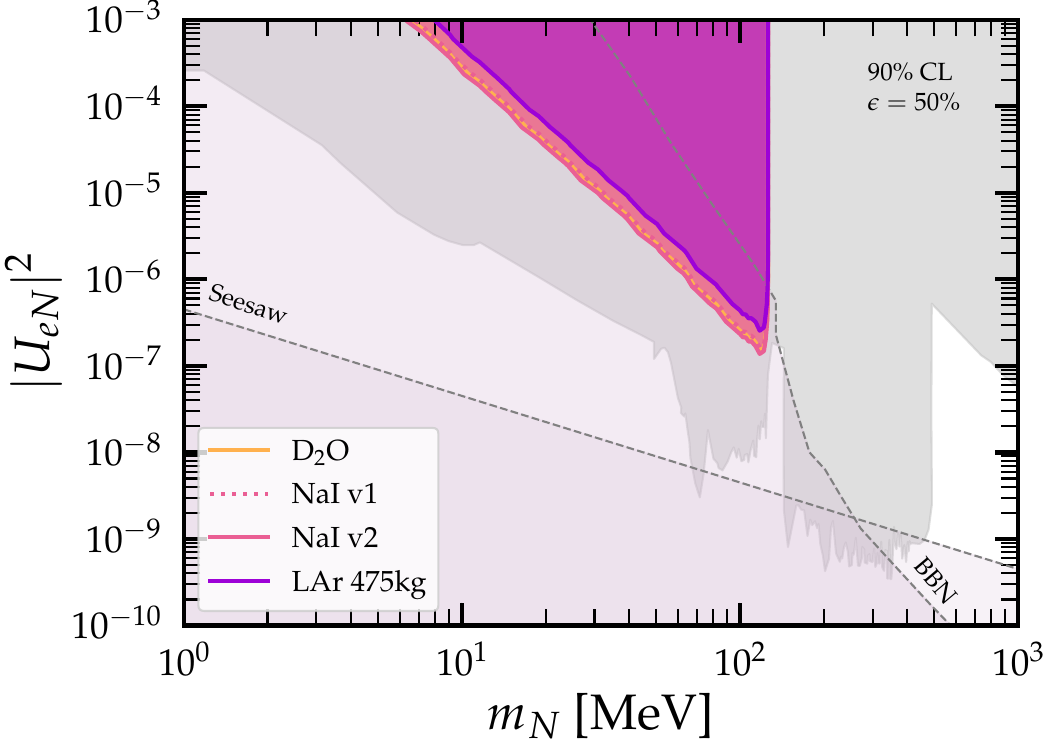}
    \includegraphics[scale=0.45]{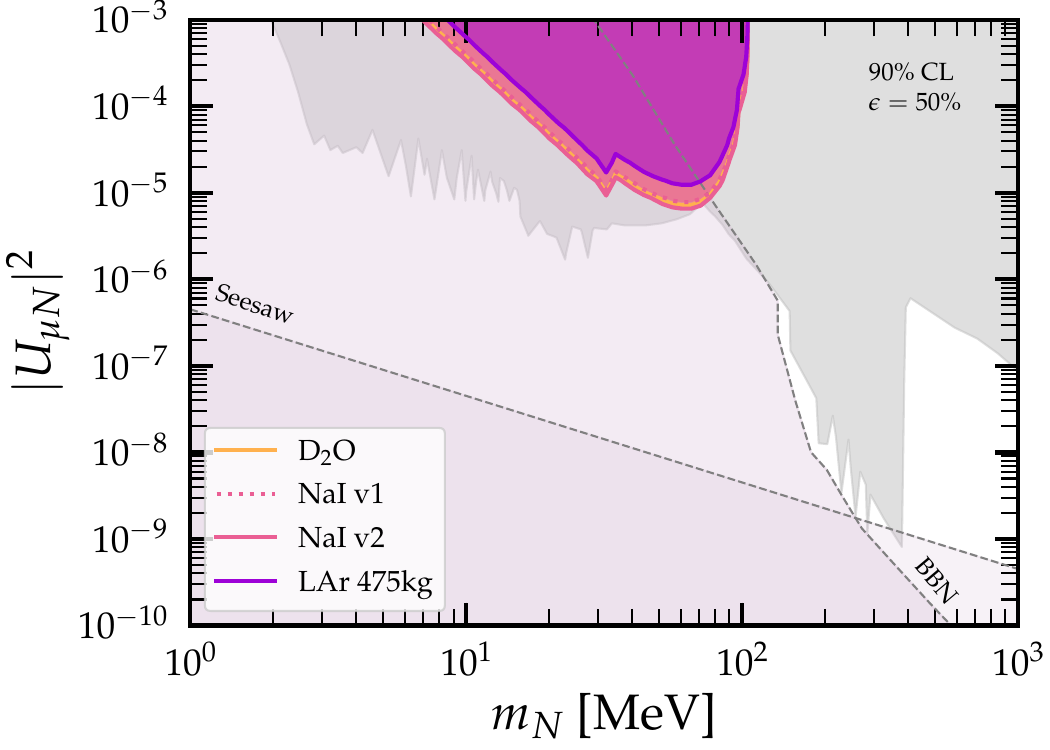}\\
    \includegraphics[scale=0.45]{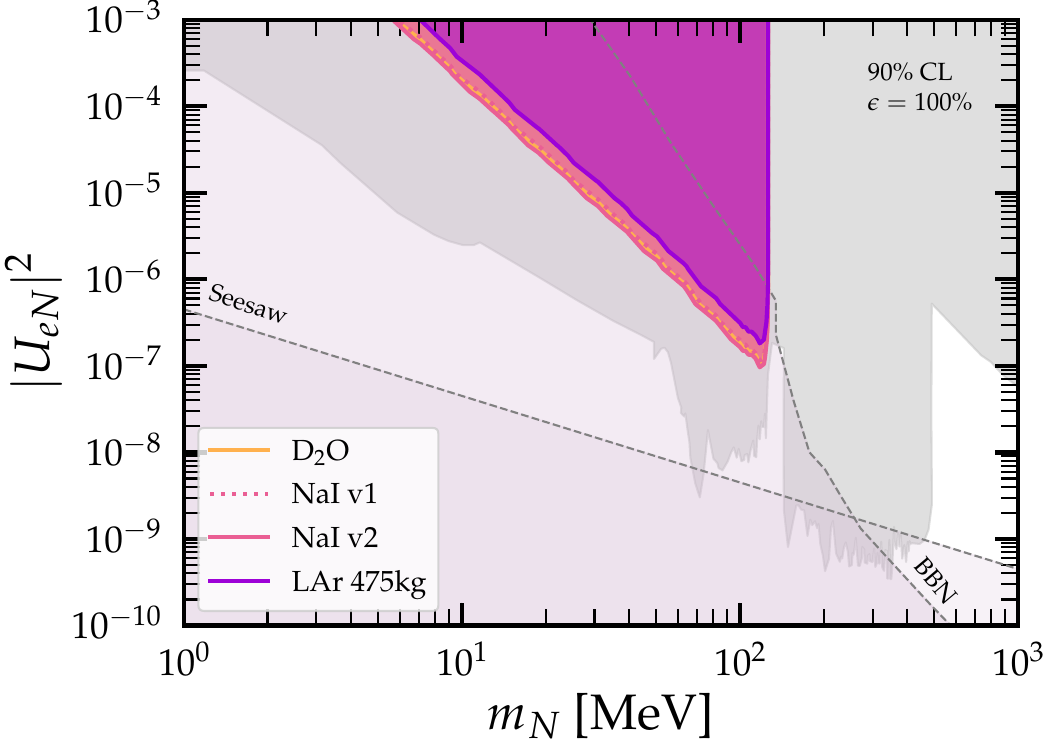}
    \includegraphics[scale=0.45]{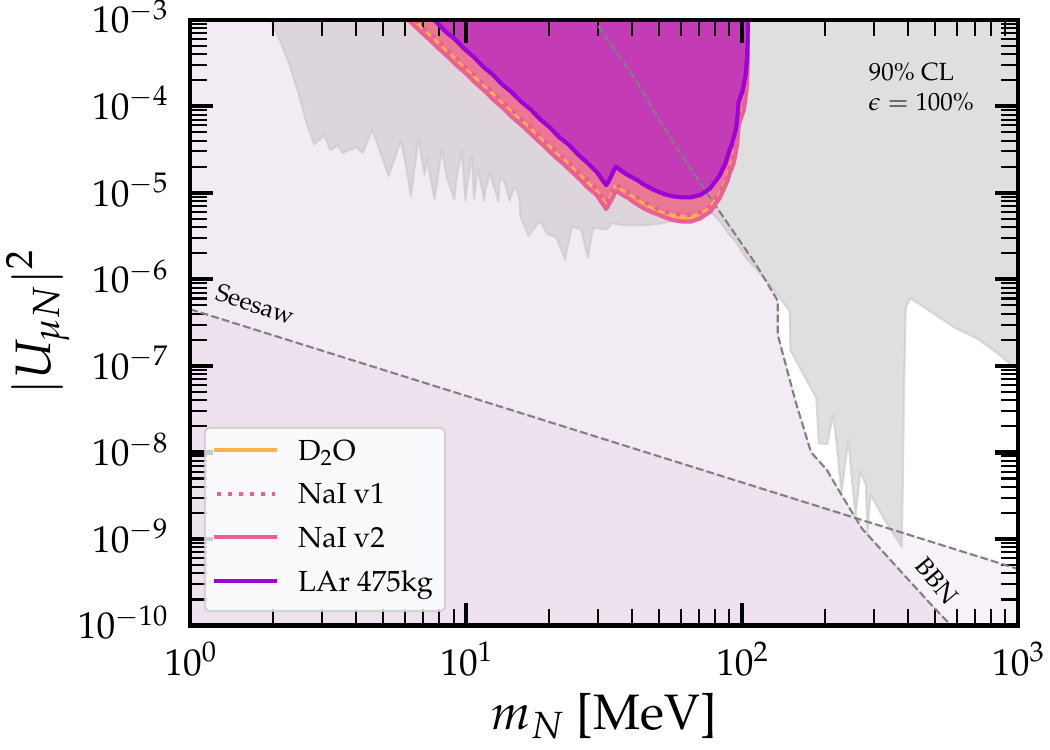}
    \caption{Same as Fig.~\ref{fig:SNS_50eff} upper panels, but assuming a total number of background events $B = 10^5$ (corresponding to $S_{\rm up} = 521.5$), with $\varepsilon = 50\%$  (\textbf{upper panels}) and $\varepsilon = 100\%$ efficiency (\textbf{lower panels}). }
    \label{fig:current_B1e5}
\end{figure*}

\begin{figure*}[htb!]
    \centering
    \includegraphics[scale=0.45]{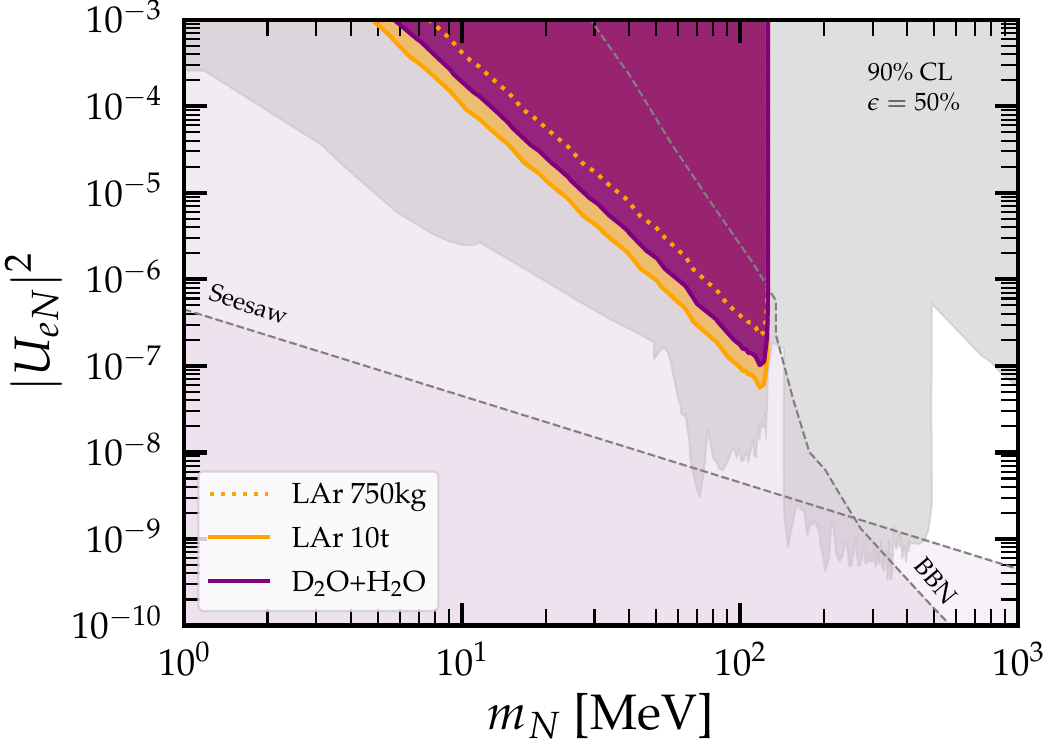}
    \includegraphics[scale=0.45]{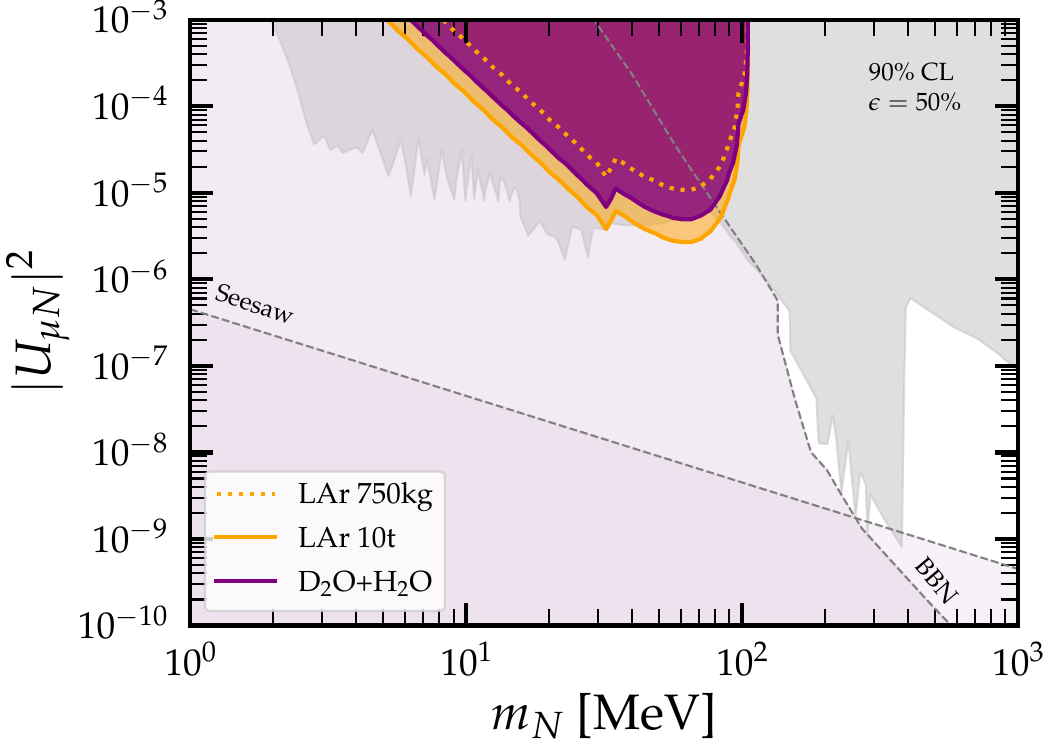}\\
    \includegraphics[scale=0.45]{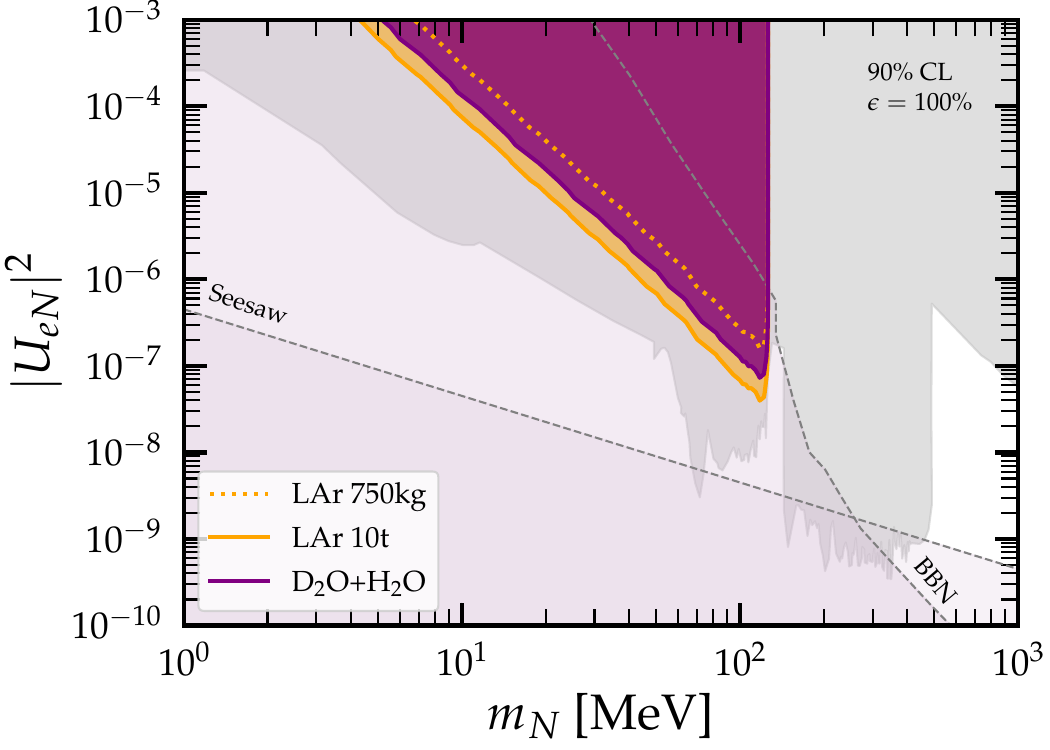}
    \includegraphics[scale=0.45]{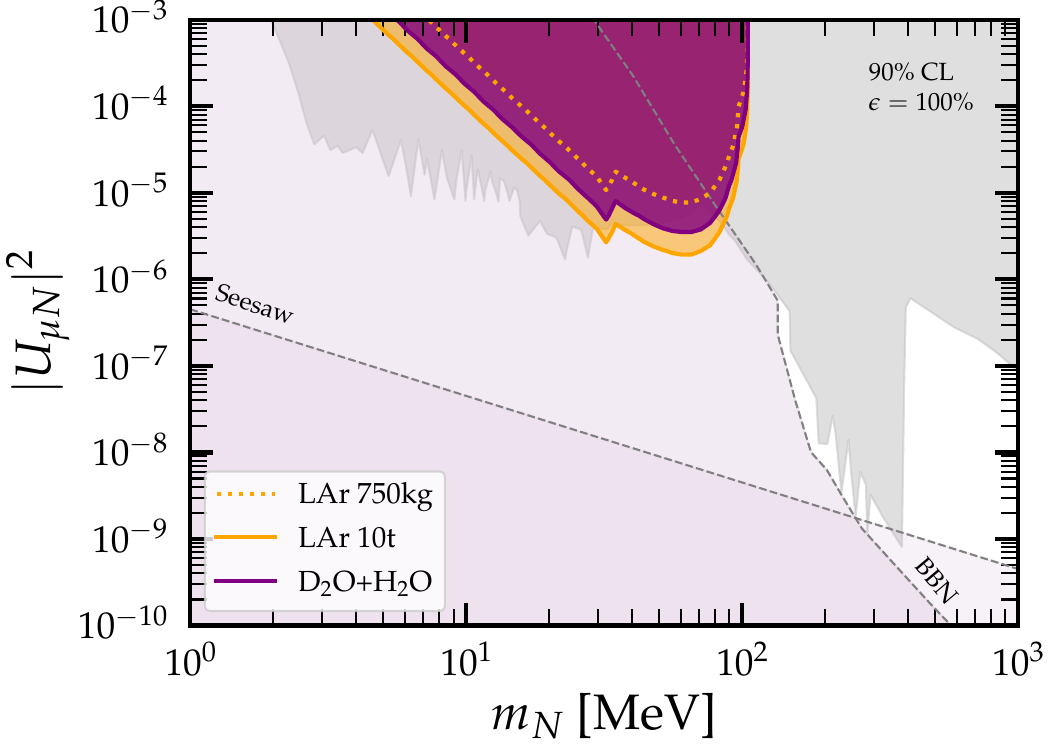}
    \caption{Same as Fig.~\ref{fig:SNS_50eff} lower panels, but assuming a total number of background events $B = 10^5$ (corresponding to $S_{\rm up} = 521.5$), with $\varepsilon = 50\%$  (\textbf{upper panels}) and $\varepsilon = 100\%$ efficiency (\textbf{lower panels}).}
    \label{fig:future_B1e5}
\end{figure*}

Similarly, we present in Figures~\ref{fig:current_B1e5} and \ref{fig:future_B1e5} the 90\% CL projected exclusion regions in the $(m_N, |U_{\ell N}|^2)$ plane, for $\ell = e, \mu$, for the currently and future SNS detectors listed in Table~\ref{tab:detectors} but assuming a very conservative hypothesis for the background events, $B = 10^5$ and for two choices of efficiency.

A comparison across all panels leads to two general observations. First, increasing the flat efficiency from $50\%$ to $100\%$ yields only a moderate improvement in the exclusion reach. Second, and more significantly, the assumed background level has a strong impact on sensitivity: in the presence of sizable backgrounds, the constraints from current SNS detectors are in most cases limited to regions already excluded by existing bounds, and do not access new parameter space. This underscores the importance of dedicated background reduction 
strategies for future HNL searches at these facilities. A careful treatment of the background -- exploiting the beam timing structure, optimized detector shielding, and improved veto systems -- has the potential to significantly reduce the background contamination and restore sensitivity to previously unexplored regions of parameter space, particularly in the muon-mixing scenario. The same considerations apply to the other spallation-source facilities discussed in the main text.

\newpage
\bibliographystyle{utphys}
\bibliography{bibliography}  

@article{Abele:2022iml,
    author = "Abele, H. and others",
    title = "{Particle Physics at the European Spallation Source}",
    eprint = "2211.10396",
    archivePrefix = "arXiv",
    primaryClass = "physics.ins-det",
    doi = "10.1016/j.physrep.2023.06.001",
    journal = "Phys. Rept.",
    volume = "1023",
    pages = "1--84",
    year = "2023"
}

@misc{CICENNS2025talk,
  author       = {Xiang Xiao},
  title        = "{Status of the CICENNS Experiment}",
  year         = {2025},
  url          = {https://indico.cern.ch/event/1458241/contributions/6522284/attachments/3082978/5457435/CICENNS_XiangXiao.pdf},
  note         = {Talk at Magnificent CE$\nu$NS 2025}
}

@article{Su:2023klh,
    author = "Su, Chenguang and Liu, Qian and Liang, Tianjiao",
    collaboration = "CLOVERS, CE{\ensuremath{\nu}}NS@CSNS",
    title = "{CE{\ensuremath{\nu}}NS Experiment Proposal at CSNS {\textdagger}}",
    eprint = "2303.13423",
    archivePrefix = "arXiv",
    primaryClass = "physics.ins-det",
    doi = "10.3390/psf2023008019",
    journal = "Phys. Sci. Forum",
    volume = "8",
    number = "1",
    pages = "19",
    year = "2023"
}

@article{JSNS2:2013jdh,
    author = "Harada, M. and others",
    collaboration = "JSNS2",
    title = "{Proposal: A Search for Sterile Neutrino at J-PARC Materials and Life Science Experimental Facility}",
    eprint = "1310.1437",
    archivePrefix = "arXiv",
    primaryClass = "physics.ins-det",
    month = "10",
    year = "2013"
}

@article{Collar:2025sle,
    author = "Collar, J. I. and others",
    title = "{Coherent elastic neutrino-nucleus scattering at the Japan Proton Accelerator Research Complex}",
    eprint = "2512.19788",
    archivePrefix = "arXiv",
    primaryClass = "hep-ph",
    doi = "10.1007/JHEP05(2026)106",
    journal = "JHEP",
    volume = "05",
    pages = "106",
    year = "2026"
}

@article{Ema:2023buz,
    author = "Ema, Yohei and Liu, Zhen and Lyu, Kun-Feng and Pospelov, Maxim",
    title = "{Heavy Neutral Leptons from Stopped Muons and Pions}",
    eprint = "2306.07315",
    archivePrefix = "arXiv",
    primaryClass = "hep-ph",
    reportNumber = "UMN-TH-4217/23, FTPI-MINN-23-10",
    doi = "10.1007/JHEP08(2023)169",
    journal = "JHEP",
    volume = "08",
    pages = "169",
    year = "2023"
}

@article{Hostert:2025ffy,
    author = "Hostert, Matheus and Urrea, Salvador",
    title = "{Long-Lived Particles from Meson and Muon Decays at Rest at Spallation Sources}",
    eprint = "2509.14085",
    archivePrefix = "arXiv",
    primaryClass = "hep-ph",
    month = "9",
    year = "2025"
}

@article{Arguelles:2021dqn,
    author = {Arg{\"u}elles, Carlos A. and Foppiani, Nicol{\`o} and Hostert, Matheus},
    title = "{Heavy neutral leptons below the kaon mass at hodoscopic neutrino detectors}",
    eprint = "2109.03831",
    archivePrefix = "arXiv",
    primaryClass = "hep-ph",
    doi = "10.1103/PhysRevD.105.095006",
    journal = "Phys. Rev. D",
    volume = "105",
    number = "9",
    pages = "095006",
    year = "2022"
}

@article{Bryman:2019bjg,
    author = "Bryman, D. A. and Shrock, R.",
    title = "{Constraints on Sterile Neutrinos in the MeV to GeV Mass Range}",
    eprint = "1909.11198",
    archivePrefix = "arXiv",
    primaryClass = "hep-ph",
    reportNumber = "TRIUMF-UBC-Stony Brook preprint (YITP-SB-2019-9)",
    doi = "10.1103/PhysRevD.100.073011",
    journal = "Phys. Rev. D",
    volume = "100",
    pages = "073011",
    year = "2019"
}

@article{Kelly:2021xbv,
    author = "Kelly, Kevin James and Machado, Pedro A. N.",
    title = "{MicroBooNE experiment, NuMI absorber, and heavy neutral leptons}",
    eprint = "2106.06548",
    archivePrefix = "arXiv",
    primaryClass = "hep-ph",
    reportNumber = "FERMILAB-PUB-21-277-T",
    doi = "10.1103/PhysRevD.104.055015",
    journal = "Phys. Rev. D",
    volume = "104",
    number = "5",
    pages = "055015",
    year = "2021"
}

@article{AristizabalSierra:2026jgp,
    author = "Aristizabal Sierra, D. and De Romeri, V. and Papoulias, D. K. and Sanchez Garcia, G.",
    title = "{Sensitivity to sub-GeV dark matter in forthcoming spallation-source neutrino experiments}",
    eprint = "2603.02132",
    archivePrefix = "arXiv",
    primaryClass = "hep-ph",
    month = "3",
    year = "2026"
}

@article{Freedman:1973yd,
    author = "Freedman, Daniel Z.",
    title = "{Coherent Neutrino Nucleus Scattering as a Probe of the Weak Neutral Current}",
    reportNumber = "NAL-PUB-73-76-THY, FERMILAB-PUB-73-076-T",
    doi = "10.1103/PhysRevD.9.1389",
    journal = "Phys. Rev. D",
    volume = "9",
    pages = "1389--1392",
    year = "1974"
}

@article{Adamski:2024yqt,
    author = "Adamski, S. and others",
    title = "{First detection of coherent elastic neutrino-nucleus scattering on germanium}",
    eprint = "2406.13806",
    archivePrefix = "arXiv",
    primaryClass = "hep-ex",
    month = "6",
    year = "2024"
}

@article{Colaresi:2022obx,
    author = "Colaresi, J. and Collar, J. I. and Hossbach, T. W. and Lewis, C. M. and Yocum, K. M.",
    title = "{Measurement of Coherent Elastic Neutrino-Nucleus Scattering from Reactor Antineutrinos}",
    eprint = "2202.09672",
    archivePrefix = "arXiv",
    primaryClass = "hep-ex",
    doi = "10.1103/PhysRevLett.129.211802",
    journal = "Phys. Rev. Lett.",
    volume = "129",
    number = "21",
    pages = "211802",
    year = "2022"
}

@article{PandaX:2024muv,
    author = "Bo, Zihao and others",
    collaboration = "PandaX",
    title = "{First Indication of Solar B8 Neutrinos through Coherent Elastic Neutrino-Nucleus Scattering in PandaX-4T}",
    eprint = "2407.10892",
    archivePrefix = "arXiv",
    primaryClass = "hep-ex",
    doi = "10.1103/PhysRevLett.133.191001",
    journal = "Phys. Rev. Lett.",
    volume = "133",
    number = "19",
    pages = "191001",
    year = "2024"
}

@article{XENON:2024ijk,
    author = "Aprile, Elena and others",
    collaboration = "XENON",
    title = "{First Indication of Solar $^8B$~Neutrinos via Coherent Elastic Neutrino-Nucleus Scattering with XENONnT}",
    eprint = "2408.02877",
    archivePrefix = "arXiv",
    primaryClass = "nucl-ex",
    doi = "10.1103/PhysRevLett.133.191002",
    journal = "Phys. Rev. Lett.",
    volume = "133",
    pages = "191002",
    year = "2024"
}

@article{Papoulias:2019lfi,
    author = "Papoulias, D. K. and Kosmas, T. S. and Sahu, R. and Kota, V. K. B. and Hota, M.",
    title = "{Constraining nuclear physics parameters with current and future COHERENT data}",
    eprint = "1903.03722",
    archivePrefix = "arXiv",
    primaryClass = "hep-ph",
    reportNumber = "IFIC/19-xxx",
    doi = "10.1016/j.physletb.2019.135133",
    journal = "Phys. Lett. B",
    volume = "800",
    pages = "135133",
    year = "2020"
}

@article{COHERENT:2017ipa,
    author = "Akimov, D. and others",
    collaboration = "COHERENT",
    title = "{Observation of Coherent Elastic Neutrino-Nucleus Scattering}",
    eprint = "1708.01294",
    archivePrefix = "arXiv",
    primaryClass = "nucl-ex",
    doi = "10.1126/science.aao0990",
    journal = "Science",
    volume = "357",
    number = "6356",
    pages = "1123--1126",
    year = "2017"
}

@article{DeRomeri:2022twg,
    author = "De Romeri, V. and Miranda, O. G. and Papoulias, D. K. and Sanchez Garcia, G. and T{\'o}rtola, M. and Valle, J. W. F.",
    title = "{Physics implications of a combined analysis of COHERENT CsI and LAr data}",
    eprint = "2211.11905",
    archivePrefix = "arXiv",
    primaryClass = "hep-ph",
    doi = "10.1007/JHEP04(2023)035",
    journal = "JHEP",
    volume = "04",
    pages = "035",
    year = "2023"
}

@article{Cadeddu:2020lky,
    author = "Cadeddu, M. and Dordei, F. and Giunti, C. and Li, Y. F. and Picciau, E. and Zhang, Y. Y.",
    title = "{Physics results from the first COHERENT observation of coherent elastic neutrino-nucleus scattering in argon and their combination with cesium-iodide data}",
    eprint = "2005.01645",
    archivePrefix = "arXiv",
    primaryClass = "hep-ph",
    doi = "10.1103/PhysRevD.102.015030",
    journal = "Phys. Rev. D",
    volume = "102",
    number = "1",
    pages = "015030",
    year = "2020"
}

@article{Abdullah:2022zue,
    author = "Abdullah, M. and others",
    title = "{Coherent elastic neutrino-nucleus scattering: Terrestrial and astrophysical applications}",
    eprint = "2203.07361",
    archivePrefix = "arXiv",
    primaryClass = "hep-ph",
    month = "3",
    year = "2022"
}

@article{Ackermann:2025obx,
    author = "Ackermann, N. and others",
    title = "{Direct observation of coherent elastic antineutrino{\textendash}nucleus scattering}",
    eprint = "2501.05206",
    archivePrefix = "arXiv",
    primaryClass = "hep-ex",
    doi = "10.1038/s41586-025-09322-2",
    journal = "Nature",
    volume = "643",
    number = "8074",
    pages = "1229--1233",
    year = "2025"
}

@article{COHERENT:2026yje,
    author = "Adhikari, M. and others",
    collaboration = "COHERENT",
    title = "{Measurement of coherent elastic neutrino nucleus scattering on germanium by COHERENT}",
    eprint = "2603.17951",
    archivePrefix = "arXiv",
    primaryClass = "hep-ex",
    month = "3",
    year = "2026"
}

@article{LZ:2025igz,
    author = "Akerib, D. S. and others",
    collaboration = "LZ",
    title = "{Searches for Light Dark Matter and Evidence of Coherent Elastic Neutrino-Nucleus Scattering of Solar Neutrinos with the LUX-ZEPLIN (LZ) Experiment}",
    eprint = "2512.08065",
    archivePrefix = "arXiv",
    primaryClass = "hep-ex",
    month = "12",
    year = "2025"
}

@article{DeRomeri:2026prc,
    author = "De Romeri, Valentina and Papoulias, Dimitrios K. and Pompa, Federica and Sanchez Garcia, Gonzalo and Ternes, Christoph A.",
    title = "{Testing light and heavy vector mediators with solar CE$ν$NS measurements}",
    eprint = "2603.00554",
    archivePrefix = "arXiv",
    primaryClass = "hep-ph",
    month = "2",
    year = "2026"
}

@article{AtzoriCorona:2025xwr,
    author = "Atzori Corona, Mattia and Cadeddu, Matteo and Cargioli, Nicola and Dordei, Francesca and Giunti, Carlo and Ternes, Christoph A.",
    title = "{Standard Model Tested with Neutrinos}",
    eprint = "2504.05272",
    archivePrefix = "arXiv",
    primaryClass = "hep-ph",
    doi = "10.1103/dplq-dvc8",
    journal = "Phys. Rev. Lett.",
    volume = "135",
    number = "23",
    pages = "231803",
    year = "2025"
}

@article{COHERENT:2020iec,
    author = "Akimov, D. and others",
    collaboration = "COHERENT",
    title = "{First Measurement of Coherent Elastic Neutrino-Nucleus Scattering on Argon}",
    eprint = "2003.10630",
    archivePrefix = "arXiv",
    primaryClass = "nucl-ex",
    doi = "10.1103/PhysRevLett.126.012002",
    journal = "Phys. Rev. Lett.",
    volume = "126",
    number = "1",
    pages = "012002",
    year = "2021"
}

@article{COHERENT:2021xmm,
    author = "Akimov, D. and others",
    collaboration = "COHERENT",
    title = "{Measurement of the Coherent Elastic Neutrino-Nucleus Scattering Cross Section on CsI by COHERENT}",
    eprint = "2110.07730",
    archivePrefix = "arXiv",
    primaryClass = "hep-ex",
    doi = "10.1103/PhysRevLett.129.081801",
    journal = "Phys. Rev. Lett.",
    volume = "129",
    number = "8",
    pages = "081801",
    year = "2022"
}

@article{Breso-Pla:2025cul,
    author = "Bres{\'o}-Pla, V{\'\i}ctor and Cruz-Alzaga, Sergio and Gonz{\'a}lez-Alonso, Mart{\'\i}n and Prakash, Suraj",
    title = "{Muon-Decay Parameters from COHERENT}",
    eprint = "2502.18175",
    archivePrefix = "arXiv",
    primaryClass = "hep-ph",
    doi = "10.1103/mlhl-v822",
    journal = "Phys. Rev. Lett.",
    volume = "135",
    number = "13",
    pages = "131802",
    year = "2025"
}

@article{Candela:2024ljb,
    author = "Candela, Pablo M. and De Romeri, Valentina and Melas, Pantelis and Papoulias, Dimitrios K. and Saoulidou, Niki",
    title = "{Up-scattering production of a sterile fermion at DUNE: complementarity with spallation source and direct detection experiments}",
    eprint = "2404.12476",
    archivePrefix = "arXiv",
    primaryClass = "hep-ph",
    doi = "10.1007/JHEP10(2024)032",
    journal = "JHEP",
    volume = "10",
    pages = "032",
    year = "2024"
}

@article{DeRomeri:2023cjt,
    author = "De Romeri, Valentina and Lozano, Victor Martin and Sanchez Garcia, G.",
    title = "{Neutrino window to scalar leptoquarks: From low energy to colliders}",
    eprint = "2307.13790",
    archivePrefix = "arXiv",
    primaryClass = "hep-ph",
    reportNumber = "IFIC/23-27",
    doi = "10.1103/PhysRevD.109.055014",
    journal = "Phys. Rev. D",
    volume = "109",
    number = "5",
    pages = "055014",
    year = "2024"
}

@article{Candela:2023rvt,
    author = "Candela, Pablo M. and De Romeri, Valentina and Papoulias, Dimitrios K.",
    title = "{COHERENT production of a dark fermion}",
    eprint = "2305.03341",
    archivePrefix = "arXiv",
    primaryClass = "hep-ph",
    doi = "10.1103/PhysRevD.108.055001",
    journal = "Phys. Rev. D",
    volume = "108",
    number = "5",
    pages = "055001",
    year = "2023"
}

@article{Baxter:2019mcx,
    author = "Baxter, D. and others",
    title = "{Coherent Elastic Neutrino-Nucleus Scattering at the European Spallation Source}",
    eprint = "1911.00762",
    archivePrefix = "arXiv",
    primaryClass = "physics.ins-det",
    reportNumber = "IFIC/19-45, YITP-SB-19-37, FERMILAB-PUB-19-612-V",
    doi = "10.1007/JHEP02(2020)123",
    journal = "JHEP",
    volume = "02",
    pages = "123",
    year = "2020"
}

@article{Lozano:2025ekx,
    author = "Lozano, V{\'\i}ctor Mart{\'\i}n and Sanchez Garcia, G. and Terrones, Adri{\'a}n",
    title = "{Neutrino nonstandard interactions: Confronting COHERENT and LHC data}",
    eprint = "2503.11766",
    archivePrefix = "arXiv",
    primaryClass = "hep-ph",
    doi = "10.1103/fsdc-zbvv",
    journal = "Phys. Rev. D",
    volume = "112",
    number = "5",
    pages = "055017",
    year = "2025"
}

@article{COHERENT:2026ewu,
    author = "Adhikari, M. and others",
    collaboration = "COHERENT",
    title = "{The COHERENT Experiment: 2026 Update}",
    eprint = "2602.15652",
    archivePrefix = "arXiv",
    primaryClass = "hep-ex",
    month = "2",
    year = "2026"
}

@article{CentellesChulia:2025jir,
    author = "Centelles Chuli{\'a}, Salvador and Lindner, Manfred and Rink, Thomas",
    title = "{Testing lepton non-unitarity with the next generation of (Germanium-based) CE$\nu$NS reactor experiments}",
    eprint = "2512.09027",
    archivePrefix = "arXiv",
    primaryClass = "hep-ph",
    month = "12",
    year = "2025"
}

@article{Miranda:2020syh,
    author = "Miranda, O. G. and Papoulias, D. K. and Sanders, O. and T{\'o}rtola, M. and Valle, J. W. F.",
    title = "{Future CEvNS experiments as probes of lepton unitarity and light-sterile neutrinos}",
    eprint = "2008.02759",
    archivePrefix = "arXiv",
    primaryClass = "hep-ph",
    doi = "10.1103/PhysRevD.102.113014",
    journal = "Phys. Rev. D",
    volume = "102",
    pages = "113014",
    year = "2020"
}

@article{Abdullahi:2022jlv,
    author = "Abdullahi, Asli M. and others",
    title = "{The present and future status of heavy neutral leptons}",
    eprint = "2203.08039",
    archivePrefix = "arXiv",
    primaryClass = "hep-ph",
    reportNumber = "FERMILAB-CONF-22-184-T-V",
    doi = "10.1088/1361-6471/ac98f9",
    journal = "J. Phys. G",
    volume = "50",
    number = "2",
    pages = "020501",
    year = "2023"
}

@article{CCM:2021leg,
    author = "Aguilar-Arevalo, A. A. and others",
    collaboration = "CCM",
    title = "{First dark matter search results from Coherent CAPTAIN-Mills}",
    eprint = "2105.14020",
    archivePrefix = "arXiv",
    primaryClass = "hep-ex",
    reportNumber = "LA-UR-21-24983",
    doi = "10.1103/PhysRevD.106.012001",
    journal = "Phys. Rev. D",
    volume = "106",
    number = "1",
    pages = "012001",
    year = "2022"
}

@article{Ballett:2019bgd,
    author = "Ballett, Peter and Boschi, Tommaso and Pascoli, Silvia",
    title = "{Heavy Neutral Leptons from low-scale seesaws at the DUNE Near Detector}",
    eprint = "1905.00284",
    archivePrefix = "arXiv",
    primaryClass = "hep-ph",
    reportNumber = "IPPP/18/76",
    doi = "10.1007/JHEP03(2020)111",
    journal = "JHEP",
    volume = "03",
    pages = "111",
    year = "2020"
}

@article{Berryman:2019dme,
    author = "Berryman, Jeffrey M. and de Gouvea, Andre and Fox, Patrick J and Kayser, Boris Jules and Kelly, Kevin James and Raaf, Jennifer Lynne",
    title = "{Searches for Decays of New Particles in the DUNE Multi-Purpose Near Detector}",
    eprint = "1912.07622",
    archivePrefix = "arXiv",
    primaryClass = "hep-ph",
    reportNumber = "FERMILAB-PUB-19-607-ND-T, NUHEP-TH/19-16",
    doi = "10.1007/JHEP02(2020)174",
    journal = "JHEP",
    volume = "02",
    pages = "174",
    year = "2020"
}

@article{Feng:2026jsi,
    author = "Feng, Ruofei and Ge, Shao-Feng and Zhang, Yongchao",
    title = "{Probing Light Dark Particles in Neutrino Scattering Experiments}",
    eprint = "2602.08314",
    archivePrefix = "arXiv",
    primaryClass = "hep-ph",
    month = "2",
    year = "2026"
}

@article{Brdar:2018qqj,
    author = "Brdar, Vedran and Rodejohann, Werner and Xu, Xun-Jie",
    title = "{Producing a new Fermion in Coherent Elastic Neutrino-Nucleus Scattering: from Neutrino Mass to Dark Matter}",
    eprint = "1810.03626",
    archivePrefix = "arXiv",
    primaryClass = "hep-ph",
    doi = "10.1007/JHEP12(2018)024",
    journal = "JHEP",
    volume = "12",
    pages = "024",
    year = "2018"
}

@article{Chang:2020jwl,
    author = "Chang, We-Fu and Liao, Jiajun",
    title = "{Constraints on light singlet fermion interactions from coherent elastic neutrino-nucleus scattering}",
    eprint = "2002.10275",
    archivePrefix = "arXiv",
    primaryClass = "hep-ph",
    doi = "10.1103/PhysRevD.102.075004",
    journal = "Phys. Rev. D",
    volume = "102",
    number = "7",
    pages = "075004",
    year = "2020"
}

@article{Abada:2022wvh,
    author = "Abada, Asmaa and Escribano, Pablo and Marcano, Xabier and Piazza, Gioacchino",
    title = "{Collider searches for heavy neutral leptons: beyond simplified scenarios}",
    eprint = "2208.13882",
    archivePrefix = "arXiv",
    primaryClass = "hep-ph",
    reportNumber = "IFT-UAM/CSIC-22-98",
    doi = "10.1140/epjc/s10052-022-11011-7",
    journal = "Eur. Phys. J. C",
    volume = "82",
    number = "11",
    pages = "1030",
    year = "2022"
}

@article{Fernandez-Martinez:2023phj,
    author = "Fern\'andez-Mart\'\i{}nez, Enrique and Gonz\'alez-L\'opez, Manuel and Hern\'andez-Garc\'\i{}a, Josu and Hostert, Matheus and L\'opez-Pav\'on, Jacobo",
    title = "{Effective portals to heavy neutral leptons}",
    eprint = "2304.06772",
    archivePrefix = "arXiv",
    primaryClass = "hep-ph",
    reportNumber = "FTUV-23-0303.1224, IFIC/23-09",
    doi = "10.1007/JHEP09(2023)001",
    journal = "JHEP",
    volume = "09",
    pages = "001",
    year = "2023"
}

@article{ATLAS:2025qbs,
    collaboration = "ATLAS",
    title = "{Common ATLAS, CMS and LHCb summary plots for Heavy Neutral Leptons (Fall 2025)}",
    reportNumber = "ATL-PHYS-PUB-2025-048",
    year = "2025"
}

@article{Cottin:2018nms,
    author = "Cottin, Giovanna and Helo, Juan Carlos and Hirsch, Martin",
    title = "{Displaced vertices as probes of sterile neutrino mixing at the LHC}",
    eprint = "1806.05191",
    archivePrefix = "arXiv",
    primaryClass = "hep-ph",
    reportNumber = "IFIC/18-25, IFIC-18-25",
    doi = "10.1103/PhysRevD.98.035012",
    journal = "Phys. Rev. D",
    volume = "98",
    number = "3",
    pages = "035012",
    year = "2018"
}

@article{Drewes:2019fou,
    author = "Drewes, Marco and Hajer, Jan",
    title = "{Heavy Neutrinos in displaced vertex searches at the LHC and HL-LHC}",
    eprint = "1903.06100",
    archivePrefix = "arXiv",
    primaryClass = "hep-ph",
    reportNumber = "CP3-19-11",
    doi = "10.1007/JHEP02(2020)070",
    journal = "JHEP",
    volume = "02",
    pages = "070",
    year = "2020"
}

@article{Bondarenko:2019tss,
    author = "Bondarenko, Kyrylo and Boyarsky, Alexey and Ovchynnikov, Maksym and Ruchayskiy, Oleg and Shchutska, Lesya",
    title = "{Probing new physics with displaced vertices: muon tracker at CMS}",
    eprint = "1903.11918",
    archivePrefix = "arXiv",
    primaryClass = "hep-ph",
    doi = "10.1103/PhysRevD.100.075015",
    journal = "Phys. Rev. D",
    volume = "100",
    number = "7",
    pages = "075015",
    year = "2019"
}

@article{SHiP:2018xqw,
    author = "Ahdida, C. and others",
    collaboration = "SHiP",
    title = "{Sensitivity of the SHiP experiment to Heavy Neutral Leptons}",
    eprint = "1811.00930",
    archivePrefix = "arXiv",
    primaryClass = "hep-ph",
    doi = "10.1007/JHEP04(2019)077",
    journal = "JHEP",
    volume = "04",
    pages = "077",
    year = "2019"
}

@article{Batell:2020vqn,
    author = "Batell, Brian and Evans, Jared A. and Gori, Stefania and Rai, Mudit",
    title = "{Dark Scalars and Heavy Neutral Leptons at DarkQuest}",
    eprint = "2008.08108",
    archivePrefix = "arXiv",
    primaryClass = "hep-ph",
    doi = "10.1007/JHEP05(2021)049",
    journal = "JHEP",
    volume = "05",
    pages = "049",
    year = "2021"
}

@article{Yanagida:1980xy,
    author = "Yanagida, Tsutomu",
    title = "{Horizontal Symmetry and Masses of Neutrinos}",
    reportNumber = "TU-80-208",
    doi = "10.1143/PTP.64.1103",
    journal = "Prog. Theor. Phys.",
    volume = "64",
    pages = "1103",
    year = "1980"
}

@article{Minkowski:1977sc,
    author = "Minkowski, Peter",
    title = "{$\mu \to e\gamma$ at a Rate of One Out of $10^{9}$ Muon Decays?}",
    reportNumber = "Print-77-0182 (BERN)",
    doi = "10.1016/0370-2693(77)90435-X",
    journal = "Phys. Lett. B",
    volume = "67",
    pages = "421--428",
    year = "1977"
}

@article{Drewes:2022akb,
    author = "Drewes, Marco and Klari{\'c}, Juraj and L{\'o}pez-Pav{\'o}n, Jacobo",
    title = "{New benchmark models for heavy neutral lepton searches}",
    eprint = "2207.02742",
    archivePrefix = "arXiv",
    primaryClass = "hep-ph",
    doi = "10.1140/epjc/s10052-022-11100-7",
    journal = "Eur. Phys. J. C",
    volume = "82",
    number = "12",
    pages = "1176",
    year = "2022"
}

@article{ParticleDataGroup:2026aaa,
    author = "Takahashi, F. and others",
    collaboration = "Particle Data Group",
    title = "{Review of Particle Physics}",
    doi = "10.1142/S0217751X26300115",
    journal = "Int. J. Mod. Phys. A",
    volume = "41",
    pages = "2630011",
    year = "2026"
}

@article{DeRomeri:2026dac,
    author = "De Romeri, Valentina and Duque, Laura and Papoulias, Dimitrios K. and Sanchez Garcia, G. and Ternes, Christoph A.",
    title = "{Refined extraction of electroweak and nuclear parameters from germanium CE$ν$NS data}",
    eprint = "2605.27121",
    archivePrefix = "arXiv",
    primaryClass = "hep-ph",
    month = "5",
    year = "2026"
}

@article{deSalas:2020pgw,
    author = "de Salas, P. F. and Forero, D. V. and Gariazzo, S. and Mart{\'\i}nez-Mirav{\'e}, P. and Mena, O. and Ternes, C. A. and T{\'o}rtola, M. and Valle, J. W. F.",
    title = "{2020 global reassessment of the neutrino oscillation picture}",
    eprint = "2006.11237",
    archivePrefix = "arXiv",
    primaryClass = "hep-ph",
    doi = "10.1007/JHEP02(2021)071",
    journal = "JHEP",
    volume = "02",
    pages = "071",
    year = "2021"
}

@article{Atre:2009rg,
    author = "Atre, Anupama and Han, Tao and Pascoli, Silvia and Zhang, Bin",
    title = "{The Search for Heavy Majorana Neutrinos}",
    eprint = "0901.3589",
    archivePrefix = "arXiv",
    primaryClass = "hep-ph",
    reportNumber = "FERMILAB-PUB-08-086-T, NSF-KITP-08-54, MADPH-06-1466, DCPT-07-198, IPPP-07-99",
    doi = "10.1088/1126-6708/2009/05/030",
    journal = "JHEP",
    volume = "05",
    pages = "030",
    year = "2009"
}

@article{Bondarenko:2018ptm,
    author = "Bondarenko, Kyrylo and Boyarsky, Alexey and Gorbunov, Dmitry and Ruchayskiy, Oleg",
    title = "{Phenomenology of GeV-scale Heavy Neutral Leptons}",
    eprint = "1805.08567",
    archivePrefix = "arXiv",
    primaryClass = "hep-ph",
    doi = "10.1007/JHEP11(2018)032",
    journal = "JHEP",
    volume = "11",
    pages = "032",
    year = "2018"
}

@article{Esteban:2024eli,
    author = "Esteban, Ivan and Gonzalez-Garcia, M. C. and Maltoni, Michele and Martinez-Soler, Ivan and Pinheiro, Jo{\~a}o Paulo and Schwetz, Thomas",
    title = "{NuFit-6.0: updated global analysis of three-flavor neutrino oscillations}",
    eprint = "2410.05380",
    archivePrefix = "arXiv",
    primaryClass = "hep-ph",
    reportNumber = "IFT-UAM/CSIC-24-140, YITP-SB-2024-24, IPPP/24/64, IPPP/24/64, IFT-UAM/CSIC-24-140, YITP-SB-2024-24",
    doi = "10.1007/JHEP12(2024)216",
    journal = "JHEP",
    volume = "12",
    pages = "216",
    year = "2024"
}

@article{Capozzi:2025wyn,
    author = "Capozzi, Francesco and Giar{\`e}, William and Lisi, Eligio and Marrone, Antonio and Melchiorri, Alessandro and Palazzo, Antonio",
    title = "{Neutrino masses and mixing: Entering the era of subpercent precision}",
    eprint = "2503.07752",
    archivePrefix = "arXiv",
    primaryClass = "hep-ph",
    doi = "10.1103/PhysRevD.111.093006",
    journal = "Phys. Rev. D",
    volume = "111",
    number = "9",
    pages = "093006",
    year = "2025"
}

@article{Mohapatra:1986bd,
    author = "Mohapatra, R. N. and Valle, J. W. F.",
    title = "{Neutrino Mass and Baryon Number Nonconservation in Superstring Models}",
    reportNumber = "MdDP-PP-86-127",
    doi = "10.1103/PhysRevD.34.1642",
    journal = "Phys. Rev. D",
    volume = "34",
    pages = "1642",
    year = "1986"
}

@article{Malinsky:2005bi,
    author = "Malinsky, Michal and Romao, J. C. and Valle, J. W. F.",
    title = "{Novel supersymmetric SO(10) seesaw mechanism}",
    eprint = "hep-ph/0506296",
    archivePrefix = "arXiv",
    reportNumber = "IFIC-05-28",
    doi = "10.1103/PhysRevLett.95.161801",
    journal = "Phys. Rev. Lett.",
    volume = "95",
    pages = "161801",
    year = "2005"
}

@article{Akhmedov:1995ip,
    author = "Akhmedov, Evgeny K. and Lindner, Manfred and Schnapka, Erhard and Valle, J. W. F.",
    title = "{Left-right symmetry breaking in NJL approach}",
    eprint = "hep-ph/9507275",
    archivePrefix = "arXiv",
    reportNumber = "IC-95-125, TUM-HEP-221-95, MPI-PHT-95-35, FTUV-95-34, IFIC-95-36",
    doi = "10.1016/0370-2693(95)01504-3",
    journal = "Phys. Lett. B",
    volume = "368",
    pages = "270--280",
    year = "1996"
}

@article{Akhmedov:1995vm,
    author = "Akhmedov, Evgeny K. and Lindner, Manfred and Schnapka, Erhard and Valle, J. W. F.",
    title = "{Dynamical left-right symmetry breaking}",
    eprint = "hep-ph/9509255",
    archivePrefix = "arXiv",
    reportNumber = "IC-95-126, TUM-HEP-222-95, MPI-PHT-95-70, FTUV-95-36, IFIC-95-38",
    doi = "10.1103/PhysRevD.53.2752",
    journal = "Phys. Rev. D",
    volume = "53",
    pages = "2752--2780",
    year = "1996"
}

@article{GONZALEZGARCIA1989360,
title = {Fast decaying neutrinos and observable flavour violation in a new class of majoron models},
journal = {Physics Letters B},
volume = {216},
number = {3},
pages = {360-366},
year = {1989},
issn = {0370-2693},
doi = {https://doi.org/10.1016/0370-2693(89)91131-3},
url = {https://www.sciencedirect.com/science/article/pii/0370269389911313},
author = {M.C. Gonzalez-Garcia and J.W.F. Valle}
}

@article{COHERENT:2021yvp,
    author = "Akimov, D. and others",
    collaboration = "COHERENT",
    title = "{Simulating the neutrino flux from the Spallation Neutron Source for the COHERENT experiment}",
    eprint = "2109.11049",
    archivePrefix = "arXiv",
    primaryClass = "hep-ex",
    doi = "10.1103/PhysRevD.106.032003",
    journal = "Phys. Rev. D",
    volume = "106",
    number = "3",
    pages = "032003",
    year = "2022"
}

@article{Britton:1992xv,
    author = "Britton, D. I. and others",
    title = "{Improved search for massive neutrinos in pi+ ---{\ensuremath{>}} e+ neutrino decay}",
    reportNumber = "TRI-PP-92-29",
    doi = "10.1103/PhysRevD.46.R885",
    journal = "Phys. Rev. D",
    volume = "46",
    pages = "R885--R887",
    year = "1992"
}

@article{PIENU:2017wbj,
    author = "Aguilar-Arevalo, A. and others",
    collaboration = "PIENU",
    title = "{Improved search for heavy neutrinos in the decay $\pi\rightarrow e\nu$}",
    eprint = "1712.03275",
    archivePrefix = "arXiv",
    primaryClass = "hep-ex",
    doi = "10.1103/PhysRevD.97.072012",
    journal = "Phys. Rev. D",
    volume = "97",
    number = "7",
    pages = "072012",
    year = "2018"
}

@article{Borexino:2013bot,
    author = "Bellini, G. and others",
    collaboration = "Borexino",
    title = "{New limits on heavy sterile neutrino mixing in B8 decay obtained with the Borexino detector}",
    eprint = "1311.5347",
    archivePrefix = "arXiv",
    primaryClass = "hep-ex",
    doi = "10.1103/PhysRevD.88.072010",
    journal = "Phys. Rev. D",
    volume = "88",
    number = "7",
    pages = "072010",
    year = "2013"
}

@article{NA62:2020mcv,
    author = "Cortina Gil, Eduardo and others",
    collaboration = "NA62",
    title = "{Search for heavy neutral lepton production in K+ decays to positrons}",
    eprint = "2005.09575",
    archivePrefix = "arXiv",
    primaryClass = "hep-ex",
    reportNumber = "CERN-EP-2020-089",
    doi = "10.1016/j.physletb.2020.135599",
    journal = "Phys. Lett. B",
    volume = "807",
    pages = "135599",
    year = "2020"
}

@article{T2K:2019jwa,
    author = "Abe, K. and others",
    collaboration = "T2K",
    title = "{Search for heavy neutrinos with the T2K near detector ND280}",
    eprint = "1902.07598",
    archivePrefix = "arXiv",
    primaryClass = "hep-ex",
    doi = "10.1103/PhysRevD.100.052006",
    journal = "Phys. Rev. D",
    volume = "100",
    number = "5",
    pages = "052006",
    year = "2019"
}

@article{CHARM:1985nku,
    author = "Bergsma, F. and others",
    collaboration = "CHARM",
    title = "{A Search for Decays of Heavy Neutrinos in the Mass Range 0.5-{GeV} to 2.8-{GeV}}",
    reportNumber = "CERN-EP-85-190",
    doi = "10.1016/0370-2693(86)91601-1",
    journal = "Phys. Lett. B",
    volume = "166",
    pages = "473--478",
    year = "1986"
}

@article{WA66:1985mfx,
    author = "Cooper-Sarkar, Amanda M. and others",
    collaboration = "WA66",
    title = "{Search for Heavy Neutrino Decays in the {BEBC} Beam Dump Experiment}",
    reportNumber = "CERN-EP-85-104",
    doi = "10.1016/0370-2693(85)91493-5",
    journal = "Phys. Lett. B",
    volume = "160",
    pages = "207--211",
    year = "1985"
}

@article{Barouki:2022bkt,
    author = "Barouki, Ryan and Marocco, Giacomo and Sarkar, Subir",
    title = "{Blast from the past II: Constraints on heavy neutral leptons from the BEBC WA66 beam dump experiment}",
    eprint = "2208.00416",
    archivePrefix = "arXiv",
    primaryClass = "hep-ph",
    doi = "10.21468/SciPostPhys.13.5.118",
    journal = "SciPost Phys.",
    volume = "13",
    pages = "118",
    year = "2022"
}

@article{Daum:1987bg,
    author = "Daum, M. and Jost, B. and Marshall, R. M. and Minehart, R. C. and Stephens, W. A. and Ziock, K. O. H.",
    title = "{Search for Admixtures of Massive Neutrinos in the Decay $\pi^+ \to \mu^+$ Neutrino}",
    reportNumber = "SIN-PR-87-04",
    doi = "10.1103/PhysRevD.36.2624",
    journal = "Phys. Rev. D",
    volume = "36",
    pages = "2624",
    year = "1987"
}

@article{PIENU:2019usb,
    author = "Aguilar-Arevalo, A. and others",
    collaboration = "PIENU",
    title = "{Search for heavy neutrinos in $\pi \to \mu\nu$ decay}",
    eprint = "1904.03269",
    archivePrefix = "arXiv",
    primaryClass = "hep-ex",
    doi = "10.1016/j.physletb.2019.134980",
    journal = "Phys. Lett. B",
    volume = "798",
    pages = "134980",
    year = "2019"
}

@article{Bernardi:1985ny,
    author = "Bernardi, G. and others",
    title = "{Search for Neutrino Decay}",
    reportNumber = "CERN-EP/85-177",
    doi = "10.1016/0370-2693(86)91602-3",
    journal = "Phys. Lett. B",
    volume = "166",
    pages = "479--483",
    year = "1986"
}

@article{Bernardi:1987ek,
    author = "Bernardi, G. and others",
    title = "{FURTHER LIMITS ON HEAVY NEUTRINO COUPLINGS}",
    reportNumber = "CERN-EP/87-234",
    doi = "10.1016/0370-2693(88)90563-1",
    journal = "Phys. Lett. B",
    volume = "203",
    pages = "332--334",
    year = "1988"
}

@article{Hayano:1982wu,
    author = "Hayano, R. S. and others",
    title = "{HEAVY NEUTRINO SEARCH USING K(mu2) DECAY}",
    reportNumber = "KEK-Preprint-82-21",
    doi = "10.1103/PhysRevLett.49.1305",
    journal = "Phys. Rev. Lett.",
    volume = "49",
    pages = "1305",
    year = "1982"
}

@article{Yamazaki:1984sj,
    author = "Yamazaki, T. and others",
    editor = "Meyer, A. and Wieczorek, E.",
    title = "{Search for Heavy Neutrinos in Kaon Decay}",
    journal = "Conf. Proc. C",
    volume = "840719",
    pages = "262",
    year = "1984"
}

@article{BNL-E949:2009dza,
    author = "Artamonov, A. V. and others",
    collaboration = "BNL-E949",
    title = "{Study of the decay $K^+\to\pi^+\nu \bar\nu$ in the momentum region $140 < P_\pi < 199$ MeV/c}",
    eprint = "0903.0030",
    archivePrefix = "arXiv",
    primaryClass = "hep-ex",
    reportNumber = "BNL-81786-2008-JA, FERMILAB-PUB-09-007-CD-T, KEK-2008-44, TRIUMF-TRI-PP-08-26, UHEP-EX-08-004",
    doi = "10.1103/PhysRevD.79.092004",
    journal = "Phys. Rev. D",
    volume = "79",
    pages = "092004",
    year = "2009"
}

@article{NA62:2021bji,
    author = "Cortina Gil, Eduardo and others",
    collaboration = "NA62",
    title = "{Search for $K^+$ decays to a muon and invisible particles}",
    eprint = "2101.12304",
    archivePrefix = "arXiv",
    primaryClass = "hep-ex",
    reportNumber = "CERN-EP-2021-018",
    doi = "10.1016/j.physletb.2021.136259",
    journal = "Phys. Lett. B",
    volume = "816",
    pages = "136259",
    year = "2021"
}

@article{NuTeV:1999kej,
    author = "Vaitaitis, A. and others",
    collaboration = "NuTeV, E815",
    title = "{Search for neutral heavy leptons in a high-energy neutrino beam}",
    eprint = "hep-ex/9908011",
    archivePrefix = "arXiv",
    reportNumber = "FERMILAB-PUB-99-223-E",
    doi = "10.1103/PhysRevLett.83.4943",
    journal = "Phys. Rev. Lett.",
    volume = "83",
    pages = "4943--4946",
    year = "1999"
}

@article{Kusenko:2009up,
    author = "Kusenko, Alexander",
    title = "{Sterile neutrinos: The Dark side of the light fermions}",
    eprint = "0906.2968",
    archivePrefix = "arXiv",
    primaryClass = "hep-ph",
    reportNumber = "UCLA-09-TEP-55",
    doi = "10.1016/j.physrep.2009.07.004",
    journal = "Phys. Rept.",
    volume = "481",
    pages = "1--28",
    year = "2009"
}

@article{Helo:2010cw,
    author = "Helo, Juan Carlos and Kovalenko, Sergey and Schmidt, Ivan",
    title = "{Sterile neutrinos in lepton number and lepton flavor violating decays}",
    eprint = "1005.1607",
    archivePrefix = "arXiv",
    primaryClass = "hep-ph",
    doi = "10.1016/j.nuclphysb.2011.07.020",
    journal = "Nucl. Phys. B",
    volume = "853",
    pages = "80--104",
    year = "2011"
}

@article{Boyarsky:2009ix,
    author = "Boyarsky, Alexey and Ruchayskiy, Oleg and Shaposhnikov, Mikhail",
    title = "{The Role of sterile neutrinos in cosmology and astrophysics}",
    eprint = "0901.0011",
    archivePrefix = "arXiv",
    primaryClass = "hep-ph",
    doi = "10.1146/annurev.nucl.010909.083654",
    journal = "Ann. Rev. Nucl. Part. Sci.",
    volume = "59",
    pages = "191--214",
    year = "2009"
}

@article{Ruchayskiy:2012si,
    author = "Ruchayskiy, Oleg and Ivashko, Artem",
    title = "{Restrictions on the lifetime of sterile neutrinos from primordial nucleosynthesis}",
    eprint = "1202.2841",
    archivePrefix = "arXiv",
    primaryClass = "hep-ph",
    reportNumber = "CERN-PH-TH-2012-042",
    doi = "10.1088/1475-7516/2012/10/014",
    journal = "JCAP",
    volume = "10",
    pages = "014",
    year = "2012"
}

@article{Vincent:2014rja,
    author = "Vincent, Aaron C. and Martinez, Enrique Fernandez and Hern{\'a}ndez, Pilar and Lattanzi, Massimiliano and Mena, Olga",
    title = "{Revisiting cosmological bounds on sterile neutrinos}",
    eprint = "1408.1956",
    archivePrefix = "arXiv",
    primaryClass = "astro-ph.CO",
    reportNumber = "IFIC-14-53, FTUAM-14-32, IFT-UAM-CSIC-14-075",
    doi = "10.1088/1475-7516/2015/04/006",
    journal = "JCAP",
    volume = "04",
    pages = "006",
    year = "2015"
}

@article{Bolton:2019pcu,
    author = "Bolton, Patrick D. and Deppisch, Frank F. and Bhupal Dev, P. S.",
    title = "{Neutrinoless double beta decay versus other probes of heavy sterile neutrinos}",
    eprint = "1912.03058",
    archivePrefix = "arXiv",
    primaryClass = "hep-ph",
    doi = "10.1007/JHEP03(2020)170",
    journal = "JHEP",
    volume = "03",
    pages = "170",
    year = "2020"
}

@article{Abazajian:2012ys,
    author = "Abazajian, K. N. and others",
    title = "{Light Sterile Neutrinos: A White Paper}",
    eprint = "1204.5379",
    archivePrefix = "arXiv",
    primaryClass = "hep-ph",
    reportNumber = "FERMILAB-PUB-12-881-PPD",
    month = "4",
    year = "2012"
}

@article{Gariazzo:2015rra,
    author = "Gariazzo, S. and Giunti, C. and Laveder, M. and Li, Y. F. and Zavanin, E. M.",
    title = "{Light sterile neutrinos}",
    eprint = "1507.08204",
    archivePrefix = "arXiv",
    primaryClass = "hep-ph",
    doi = "10.1088/0954-3899/43/3/033001",
    journal = "J. Phys. G",
    volume = "43",
    pages = "033001",
    year = "2016"
}

@article{Dutta:2020vop,
    author = "Dutta, Bhaskar and Kim, Doojin and Liao, Shu and Park, Jong-Chul and Shin, Seodong and Strigari, Louis E. and Thompson, Adrian",
    title = "{Searching for dark matter signals in timing spectra at neutrino experiments}",
    eprint = "2006.09386",
    archivePrefix = "arXiv",
    primaryClass = "hep-ph",
    reportNumber = "MI-TH-2014",
    doi = "10.1007/JHEP01(2022)144",
    journal = "JHEP",
    volume = "01",
    pages = "144",
    year = "2022"
}

@article{Dutta:2019nbn,
    author = "Dutta, Bhaskar and Kim, Doojin and Liao, Shu and Park, Jong-Chul and Shin, Seodong and Strigari, Louis E.",
    title = "{Dark matter signals from timing spectra at neutrino experiments}",
    eprint = "1906.10745",
    archivePrefix = "arXiv",
    primaryClass = "hep-ph",
    reportNumber = "MI-TH-1925",
    doi = "10.1103/PhysRevLett.124.121802",
    journal = "Phys. Rev. Lett.",
    volume = "124",
    number = "12",
    pages = "121802",
    year = "2020"
}

@article{deNiverville:2015mwa,
    author = "deNiverville, Patrick and Pospelov, Maxim and Ritz, Adam",
    title = "{Light new physics in coherent neutrino-nucleus scattering experiments}",
    eprint = "1505.07805",
    archivePrefix = "arXiv",
    primaryClass = "hep-ph",
    doi = "10.1103/PhysRevD.92.095005",
    journal = "Phys. Rev. D",
    volume = "92",
    number = "9",
    pages = "095005",
    year = "2015"
}

@article{Ge:2017mcq,
    author = "Ge, Shao-Feng and Shoemaker, Ian M.",
    title = "{Constraining Photon Portal Dark Matter with Texono and Coherent Data}",
    eprint = "1710.10889",
    archivePrefix = "arXiv",
    primaryClass = "hep-ph",
    reportNumber = "IPMU17-0149",
    doi = "10.1007/JHEP11(2018)066",
    journal = "JHEP",
    volume = "11",
    pages = "066",
    year = "2018"
}

@article{Calabrese:2022mnp,
    author = "Calabrese, Roberta and Gunn, Jacob and Miele, Gennaro and Morisi, Stefano and Roy, Samiran and Santorelli, Pietro",
    title = "{Constraining scalar leptoquarks using COHERENT data}",
    eprint = "2212.11210",
    archivePrefix = "arXiv",
    primaryClass = "hep-ph",
    doi = "10.1103/PhysRevD.107.055039",
    journal = "Phys. Rev. D",
    volume = "107",
    number = "5",
    pages = "055039",
    year = "2023"
}

@article{PROSPECT:2026jsl,
    author = "Andriamirado, M. and others",
    collaboration = "PROSPECT",
    title = "{Probing Long-Lived Particle Production in Muon Decays at the SNS with a Highly Capable Hydrocarbon Detector}",
    eprint = "2606.19299",
    archivePrefix = "arXiv",
    primaryClass = "hep-ex",
    month = "6",
    year = "2026"
}

@article{COHERENT:2021pvd,
    author = "Akimov, D. and others",
    collaboration = "COHERENT",
    title = "{First Probe of Sub-GeV Dark Matter beyond the Cosmological Expectation with the COHERENT CsI Detector at the SNS}",
    eprint = "2110.11453",
    archivePrefix = "arXiv",
    primaryClass = "hep-ex",
    doi = "10.1103/PhysRevLett.130.051803",
    journal = "Phys. Rev. Lett.",
    volume = "130",
    number = "5",
    pages = "051803",
    year = "2023"
}

@article{GEANT4:2002zbu,
    author = "Agostinelli, S. and others",
    collaboration = "GEANT4",
    title = "{GEANT4 - A Simulation Toolkit}",
    reportNumber = "SLAC-PUB-9350, FERMILAB-PUB-03-339, CERN-IT-2002-003",
    doi = "10.1016/S0168-9002(03)01368-8",
    journal = "Nucl. Instrum. Meth. A",
    volume = "506",
    pages = "250--303",
    year = "2003"
}

@article{AtzoriCorona:2026wbu,
    author = "Atzori Corona, M. and Cadeddu, M. and Cargioli, N. and Cerulli, R. and Co', G. and Dordei, F. and Giunti, C. and Pavarani, R.",
    title = "{Phenomenological implications of the high-precision COHERENT germanium CE$ν$NS data}",
    eprint = "2605.07975",
    archivePrefix = "arXiv",
    primaryClass = "hep-ph",
    month = "5",
    year = "2026"
}

@article{COHERENT:2019kwz,
    author = "Akimov, D. and others",
    collaboration = "COHERENT",
    title = "{Sensitivity of the COHERENT Experiment to Accelerator-Produced Dark Matter}",
    eprint = "1911.06422",
    archivePrefix = "arXiv",
    primaryClass = "hep-ex",
    doi = "10.1103/PhysRevD.102.052007",
    journal = "Phys. Rev. D",
    volume = "102",
    number = "5",
    pages = "052007",
    year = "2020"
}

@article{COHERENT:2023sol,
    author = "Barbeau, P. S. and others",
    collaboration = "COHERENT",
    title = "{Accessing new physics with an undoped, cryogenic CsI CEvNS detector for COHERENT at the SNS}",
    eprint = "2311.13032",
    archivePrefix = "arXiv",
    primaryClass = "hep-ex",
    doi = "10.1103/PhysRevD.109.092005",
    journal = "Phys. Rev. D",
    volume = "109",
    number = "9",
    pages = "092005",
    year = "2024"
}

@article{CCM:2021yzc,
    author = "Aguilar-Arevalo, A. A. and others",
    collaboration = "CCM",
    title = "{First Leptophobic Dark Matter Search from the Coherent{\textendash}CAPTAIN-Mills Liquid Argon Detector}",
    eprint = "2109.14146",
    archivePrefix = "arXiv",
    primaryClass = "hep-ex",
    reportNumber = "Report-no: LA-UR-21-28552, LA-UR-21-28552",
    doi = "10.1103/PhysRevLett.129.021801",
    journal = "Phys. Rev. Lett.",
    volume = "129",
    number = "2",
    pages = "021801",
    year = "2022"
}

@article{COHERENT:2022pli,
    author = "Akimov, D. Yu. and others",
    collaboration = "COHERENT",
    title = "{COHERENT constraint on leptophobic dark matter using CsI data}",
    eprint = "2205.12414",
    archivePrefix = "arXiv",
    primaryClass = "hep-ex",
    reportNumber = "FERMILAB-CONF-22-693-ND",
    doi = "10.1103/PhysRevD.106.052004",
    journal = "Phys. Rev. D",
    volume = "106",
    number = "5",
    pages = "052004",
    year = "2022"
}

@article{XENON:2026ydt,
    author = "Aprile, E. and others",
    collaboration = "XENON",
    title = "{Probing the Solar $^8$B Neutrino Fog with XENONnT}",
    eprint = "2604.06002",
    archivePrefix = "arXiv",
    primaryClass = "hep-ex",
    month = "4",
    year = "2026"
}

@article{Schechter:1980gr,
    author = "Schechter, J. and Valle, J. W. F.",
    title = "{Neutrino Masses in SU(2) x U(1) Theories}",
    reportNumber = "SU-4217-167, COO-3533-167",
    doi = "10.1103/PhysRevD.22.2227",
    journal = "Phys. Rev. D",
    volume = "22",
    pages = "2227",
    year = "1980"
}

@article{Haines:2014kna,
    author = "Haines, J. R. and McManamy, T. J. and Gabriel, T. A. and Battle, R. E. and Chipley, K. K. and Crabtree, J. A. and Jacobs, L. L. and Lousteau, D. C. and Rennich, M. J. and Riemer, B. W.",
    title = "{Spallation neutron source target station design, development, and commissioning}",
    doi = "10.1016/j.nima.2014.03.068",
    journal = "Nucl. Instrum. Meth. A",
    volume = "764",
    pages = "94--115",
    year = "2014"
}

@article{Cherkashyna:2014,
  author         = {Cherkashyna, N. and others},
  title          = {High energy particle background at neutron spallation sources and possible solutions},
  journal        = {Journal of Physics: Conference Series},
  volume         = {528},
  pages          = {012013},
  year           = {2014},
  doi            = {10.1088/1742-6596/528/1/012013}
}

@article{Bogdanova:2006ex,
    author = "Bogdanova, L. N. and Gavrilov, M. G. and Kornoukhov, V. N. and Starostin, A. S.",
    title = "{Cosmic muon flux at shallow depths underground}",
    eprint = "nucl-ex/0601019",
    archivePrefix = "arXiv",
    doi = "10.1134/S1063778806080047",
    journal = "Phys. Atom. Nucl.",
    volume = "69",
    pages = "1293--1298",
    year = "2006"
}

@article{COHERENT:2021xhx,
    author = "Akimov, D. and others",
    collaboration = "COHERENT",
    title = "{A D$_2$O detector for flux normalization of a pion decay-at-rest neutrino source}",
    eprint = "2104.09605",
    archivePrefix = "arXiv",
    primaryClass = "physics.ins-det",
    reportNumber = "FERMILAB-PUB-21-208-ND",
    doi = "10.1088/1748-0221/16/08/P08048",
    journal = "JINST",
    volume = "16",
    number = "08",
    pages = "P08048",
    year = "2021"
}

@article{Chatterjee:2022mmu,
    author = "Chatterjee, Sabya Sachi and Lavignac, St{\'e}phane and Miranda, O. G. and Sanchez Garcia, G.",
    title = "{Constraining nonstandard interactions with coherent elastic neutrino-nucleus scattering at the European Spallation Source}",
    eprint = "2208.11771",
    archivePrefix = "arXiv",
    primaryClass = "hep-ph",
    doi = "10.1103/PhysRevD.107.055019",
    journal = "Phys. Rev. D",
    volume = "107",
    number = "5",
    pages = "055019",
    year = "2023"
}

@misc{Monrabal:M7s,
title = "CEvNS at the ESS (Talk at Magnificent CEvNS)",
author = "F. Monrabal",
howpublished = {\url{ "https://indico.global/event/6083/contributions/50003/"}},
year = "2024"
}

\end{document}